%% file: main.tex
\documentclass[letterpaper,12pt]{article} 
\pdfoutput=1
\usepackage[]{algorithm2e}
\usepackage{amsmath}
\allowdisplaybreaks
\usepackage[usenames,dvipsnames]{color}
\usepackage[hmargin=0.7in,vmargin=1in]{geometry}
\usepackage[comma,longnamesfirst]{natbib}
\usepackage{booktabs}
\usepackage{amsfonts} 
\usepackage{float}
\usepackage{amsmath, amssymb}
\usepackage{graphicx}
\usepackage{epstopdf}
\usepackage{caption}
\usepackage{subcaption}
\usepackage{blindtext}
\usepackage[utf8]{inputenc}
\usepackage{setspace}
\usepackage{pdflscape}
\usepackage{numprint} 
\newtheorem{theorem}{Theorem}
\newtheorem{lemma}{Lemma}

\usepackage{fontenc}
\usepackage{epsfig}
\usepackage{rotating}
\usepackage{hyperref}
\usepackage{multirow}
\usepackage{bm}
\usepackage{algpseudocode}
\usepackage{natbib}

\newcommand\ceil[1]{\lceil#1\rceil}

\epsfverbosetrue
\usepackage{enumitem}
\newlist{props}{enumerate}{1}
\setlist[props,1]{
	label={\arabic*.},
	leftmargin=*,
	align=left,
	labelsep=5pt,
}

\makeatletter
\def\@seccntformat#1{\@ifundefined{#1@cntformat}%
   {\csname the#1\endcsname\quad}  
   {\csname #1@cntformat\endcsname}
}
\let\oldappendix\appendix 
\renewcommand\appendix{%
    \oldappendix
    \newcommand{\section@cntformat}{\appendixname~\thesection\quad}
}
\makeatother

\begin{document}
\pagestyle{empty}
\doublespacing
\input{title}

\newpage
\pagestyle{plain}
\newpage
\input{sect1}
\input{sect2}
\input{sect3}
\input{sect4}
\input{append}
\onehalfspacing

\newpage
\input{tabs}

\input{figs}

\newpage
\pagestyle{empty}
\doublespacing
\input{web-append}

\end{document}

%% file: title.tex
\begin{titlepage}
\title{{\sf Time Series Copulas for Heteroskedastic Data}}
\author{Rubén Loaiza-Maya, Michael S. Smith and Worapree Maneesoonthorn}
\date{First Version March 2016\\ This Version January 2017}
\maketitle
\noindent {\small Rub\'{e}n Loaiza-Maya is a PhD student, 
Michael Smith is Chair of Management (Econometrics) and 
Worapree Maneesoonthorn is Assistant Professor of Statistics and Econometrics, all
at Melbourne Business School, University of Melbourne.
Correspondence should be directed to Michael Smith at
{\tt mike.smith@mbs.edu}.
We thank the editor Prof. Andrew 
Patton and two anonymous referees for corrections and
constructive comments that have
improved the paper. We would also like to thank participants at the 
2016 Melbourne Bayesian Econometrics Workshop, and the 10th International Conference
on Computational
and Financial Econometrics in Seville, for useful feedback.
This
work was partially supported by
Australian Research Council Future Fellowship FT110100729.}

\newpage
\begin{center}
{\LARGE {\sf  Time Series Copulas for Heteroskedastic Data}}\\
\vspace{15pt}
{\bf Abstract}
\end{center}
\vspace{-10pt}
\noindent
We propose parametric copulas that capture serial dependence in stationary heteroskedastic time series. 
We develop our copula for first order Markov series, and extend it to higher orders and multivariate series.
We derive the copula of a volatility proxy, based on which we propose new measures of volatility dependence,
including co-movement and spillover in multivariate series. In general, these depend upon the marginal distributions 
of the series. Using exchange rate returns, we show that the resulting copula models can capture their marginal 
distributions more accurately than univariate and multivariate GARCH models, and produce more accurate value 
at risk forecasts.
\vspace{3cm}


\end{titlepage}

%% file: sect1.tex
\section{Introduction}
While parametric
copulas are widely used to model cross-sectional dependence in multivariate
time series~\citep{patton2012}, they are also increasingly employed 
to capture serial dependence in time series. We refer to the latter as
`time series copulas'. \citet{darsow1992} and \citet{ibragimov2009}
provide characterizations of time series copulas for univariate
Markov processes, while \citet{joe1997},
\citet{Lambert2002}, \citet{chen2006}, \citet{Domma2009}, \citet{chen2009}, \citet{beare2010} and
\citet{beare2012}
use Archimedean or elliptical copulas to capture
serial dependence in this case. 
\citet{SmithMinAlmCza2010} use vine copulas to capture serial dependence in 
non-stationary longitudinal data.
For multivariate time series, 
\citet{biller2003}, \citet{remillard2012},
\citet{Smith2015} and~\citet{beare15} use elliptical, Archimedean or vine copulas to account for
serial dependence. However,
all these copulas prove inadequate 
when
a time series exhibits heteroskedasticity. For example,
\citet{smithvahey2015} fit a Gaussian time series copula model to heteroskedastic multivariate 
time series data, 
but note that it has limited ability to represent serial dependence in the conditional variance.
To address this problem,
we propose 
a family of closed form parametric copulas to capture serial dependence
in heteroskedastic series. 
Using these, we construct new time series models for 
heteroskedastic continuous-valued data that also allow for flexible margins--- something
that is difficult
to achieve using existing nonlinear time series models.

Heteroskedasticity is a key feature of many financial and economic time
series.
In the multivariate case, many
authors follow \citet{Patton2006} and employ existing univariate
time series models for each series, along with a copula to account
for conditional cross-sectional dependence only. Most recently, focus has been
on dynamic specifications of the
copula parameters; 
see \citet{almeida2012}, \citet{hafner2012}, \citet{oh2013}, 
\citet{salvatierra2015} and \citet{creal2015} for some recent examples. 
\citet{SmithMan2016} consider extracting implicit or `inversion' copulas from 
univariate state space models numerically. However,
as far as we are aware, closed form
copulas that can adequately account for serial
dependence of heteroskedastic data have yet to be identified. 
To do so, we compute empirically
the bivariate copula density of first order serial dependence
for
two popular stationary heteroskedastic time series models. Both densities
have an unusual cross shape, with mass concentrated at all four corners of the unit square.
The level of concentration increases with the level of volatility persistence. We 
approximate these copulas using a mixture of bivariate
copulas.
When combined
with a flexible marginal distribution,
the resulting copula model can be employed to model
a wide range of heteroskedastic time series with Markov order one.
To illustrate, Figure~\ref{Fig:Persistent_processes} plots four simulated
time
series. Each series has
the same first order serial dependence structure,
given by
a mixture of bivariate copulas that we discuss later in Section~\ref{sec:2.1},
but with four different margins: (a)~Gaussian, (b)~t, (c)~Beta and (d)~Log-normal. Each series exhibits common
features of heteroskedastic data,
such as volatility clustering, even 
when the margin is bounded or skewed.

We extend our copula to higher Markov orders $p>1$ using a drawable vine (or `D-vine') copula.
These are compositions of bivariate copula components called `pair-copulas' \citep{Aas2009182}. 
Vine copulas can be difficult to use in high dimensions because
the number of pair-copulas and possible decompositions can be large.
However, even though the dimension is high in the time series case,
there is only one D-vine decomposition, which is parsimonious when the series
is Markov and stationary.
\citet{beare15}, \citet{brechmann15} and
\citet{Smith2015} all show that parsimonious vine copulas
can also be used to capture cross-sectional and serial dependence
jointly in multivariate time series. We follow these authors and
employ a D-vine copula for
multivariate heteroskedastic
data, but with pair-copulas given by our proposed bivariate mixture copula. 

Existing popular
dependence measures computed from the time series copula
are poor measures of volatility dependence.
A major contribution of the paper is that we derive new alternative measures.
To obtain these we consider a 
volatility proxy that is a transformation of the series,
and derive the bivariate copula of the proxy at any two points in time.
We label this a `volatility copula', and show that it is
invariant to specific choice of transformation, given some broad properties 
that are
consistent with a volatility proxy.
Then pairwise dependence measures---
such as Spearman's rho or Kendall's tau---
computed from this volatility copula can be used to
measure volatility dependence.  These pairwise measures can also 
be computed
in the multivariate time series case, forming new measures of 
volatility persistence, co-movement and spillover. The
proposed measures of volatility
dependence 
are general for two reasons. First, they are not based on a 
specific structural assumption for 
the conditional variance of the series, as is the case with most
existing
models such as the BEKK~\citep{engle1995} and DCC~\citep{engle2002} models. 
Second, they can be computed for any time series model, so that
the degree and type of volatility dependence of different
models can be compared.
To the best of our knowledge, ours is the first study to propose measuring
volatility
dependence
from the copula perspective.

The density of our bivariate mixture copula is available in closed form,
so that the model likelihoods are also. We outline parallel algorithms
to compute these efficiently for the 
vine copulas. These are extensions of that
originally proposed by~\citet{Aas2009182} to exploit the parsimonious structure
of the vine copulas in the time series case.
Maximum likelihood estimation (MLE) is 
straightforward for univariate
series with low Markov orders, but for larger vines
we follow
\citet{MinCzado2010}, \citet{SmithMinAlmCza2010} and~\citet{Smith2015}, and
compute the posterior distribution using
Markov chain Monte Carlo (MCMC) methods.

To illustrate the advantages of our new methodology we apply it to daily
foreign exchange returns. These exhibit strong 
heteroskedasticity, but have marginal distributions that are typically
asymmetric
and fat-tailed \citep{boothe1987}. Capturing such nuanced
margins is difficult using existing
time series models, but is easy in the copula framework.
We first employ a univariate time series copula model for
USD/AUD returns, and compare it to GARCH alternatives.
We then extend the study 
to also include
USD/EUR and USD/JPY returns in a trivariate time series copula model,
and 
compare it to multivariate GARCH alternatives. The GARCH models are
shown to have inaccurate margins, whereas our copula models employ more
accurate nonparametric estimates.
We compute our new metrics of volatility dependence
for all models, and find the
copula
models capture positive volatility persistence similar to the benchmark
models. 
However, in the multivariate case the copula model also captures both
positive volatility
co-movements and spillovers, whereas those from the multivariate 
GARCH models are restricted.
In a validation study we find that
the one day ahead Value-at-Risk (VaR) forecasts from the copula
models are more accurate than those from the GARCH models --- both in the univariate
and multivariate cases. A small simulation study also shows that our copulas
are more
robust to model misspecification than GARCH equivalents.

The paper is organized as follows. In Section~\ref{section:2} we outline the 
proposed copula model.
We derive the volatility copula, and show how to use it to measure
volatility persistence. The section
concludes with the analysis of USD/AUD exchange rate returns, validation
and simulation studies. 
Section~\ref{section:3} extends the methodology to 
multivariate time series, and is employed to model jointly
the three exchange rate returns series, 
while Section~\ref{section:5} concludes.

%% file: sect2.tex
\vspace{-15pt}
\section{Heteroskedastic Time Series}\label{section:2}
\vspace{-10pt}
\subsection{Copulas of Serial Dependence}\label{sec:2.1}
\vspace{-5pt}
Following \citet{Sklar1959}, the joint distribution function
of $T$ observations
$\boldsymbol{y} = (y_1,\dots,y_T)$ 
on a time series can be written as
\begin{equation}\label{eq:copcdf}
F(\bm{y})=C(\bm{u})\,.
\end{equation}
Here, $\bm{u}=(u_1,\ldots,u_T)$, $u_t=F_t(y_t)$, $F_t$ is the marginal
distribution function of $y_t$, and $C$ is a $T$-dimensional copula
function that captures the serial dependence in the time series.  
Copula functions are usually selected from a 
range of parametric copulas when modeling cross-sectional 
dependence; see, for example, \citet{Nelsen2006} and \citet{Joe2014}.
However, 
only limited consideration has been given
to an appropriate choice of
$C$ when modeling
serial dependence. We consider this here when
the time series is
heteroskedastic.

If the time series is continuous, then the density of $\bm{y}$ is
\begin{equation}\label{Eq:8} 
f(\bm{y})=c(\bm{u})\prod_{t=1}^{T}f_t(y_t)\,,
\end{equation}
where 
$f_t(y_t)=\frac{\partial}{\partial y_t}F_t(y_t)$, and
$c(\bm{u})=\frac{\partial^T}{\partial u_1,\ldots,\partial u_T}C(\bm{u})$ is the 
copula density. Note that throughout this paper, copula functions
are denoted
with upper case $C$, and copula densities with lower case $c$.
If the time series $\{y_t\}$ is strongly
stationary (\citealp{Brockwell1991}) and Markov order one, 
then it is straightforward to show that the series $\{u_t\}$ is also
\citep{Smith2015}.
In this case,
the copula density can be greatly simplified
as 
\begin{equation}\label{eq:coplike}
c(\bm{u})=\prod_{t=2}^T f(u_t|u_{t-1})= \prod_{t=2}^T c_2(u_{t-1},u_t)\,,
\end{equation}
so that the serial dependence is captured by a single bivariate
copula with density $c_2$. (Note that we choose this notation for the copula
to be
consistent with that used later for the vine copula at
Equation~(\ref{eq:DvineLikeli}).)

To explore the shape of $c_2$ for
conditionally heteroskedastic time series, we consider
two popular Markovian processes. The first is the ARCH(1) model,
where $y_t = \epsilon_t\sigma_t$, $\sigma_t^2 = \alpha_0+\alpha_1 y_{t-1}^2$,
and $\epsilon_t\sim N(0,1)$. The second is
the first order stochastic volatility model (SV(1)), where
$y_t = \epsilon_t \exp(\frac{h_t}{2})$, $(h_t-\bar h) = \phi_1 (h_{t-1}-\bar h)+\eta_t$,
and $\eta_t\sim N(0,\sigma^2)$. 
Figure~\ref{Fig:Mixture_ofCop_densities}(a,b) displays empirical copula density estimates of
$c_2$ for the ARCH(1) model with medium ($\alpha_0=0.01, \alpha_1 = 0.5$), 
and high ($\alpha_0=0.01, \alpha_1 = 0.9$) persistence. These are obtained
by simulating $T=50,000$ observations from each process, estimating
the time-invariant margins of $y_t$ using a locally adaptive kernel density 
estimator, 
from which copula data are computed. Each panel then
displays a bivariate
histogram of the copula data and their values lagged one period.
Both time series show positive and equally-valued tail dependence in all four quadrants,
along with a shallow mode
around $(0.5,0.5)$. Higher persistence results in
higher tail dependence, along with a more pronounced central mode in $c_2$.
Similar features can also be seen 
in
Figure~\ref{Fig:Mixture_ofCop_densities}(c,d), which displays the empirical copula density estimates
of $c_2$ 
for the SV(1) model with
medium ($\bar h=0.8, \sigma^2=2.5, \phi_1=0.5$) and high ($\bar h=0.8, \sigma^2=2, \phi_1=0.9$) persistence.
We note that despite the strong serial dependence in these series,
Kendall's tau and Spearman's rho--- the two most commonly employed
measures of dependence--- of $c_2$ can be shown to be exactly zero.

While most existing bivariate
parametric copulas
cannot replicate the features found
in Figure~\ref{Fig:Mixture_ofCop_densities}(a--d), mixtures of rotated copulas can do so.
Mixtures of rotated or other copulas are a popular way to produce more flexible
copulas; for example, see~\citet{fortin2002}, \citet{Smith2015}
and~\citet{oh2016} among others. 
Let $C^a,C^b$ and $c^a,c^b$ be copula
functions and densities of two parametric bivariate
copulas
that both have non-negative Kendall's tau.
(We label these copulas using superscripts to avoid confusion with
pair-copulas indices employed later.)
Then, 
we use a mixture of $c^a$ and a 90 degree rotation of $c^b$, with density
\begin{equation}
c^{MIX}(u,v;\bm{\gamma}) = wc^a(u,v;\bm{\gamma}^a)+(1-w)c^b(1-u,v;\bm{\gamma}^b)
\,,\;0<w<1\,,
\label{eq:cmix}
\end{equation}
and parameters $\bm{\gamma}=\{w,\bm{\gamma}^a,\bm{\gamma}^b\}$, for
$c_2$ in Equation~(\ref{eq:coplike}). 
For example, t copulas~\citep{demarta2005t} with positive correlation parameters
can be used
for $c^a$ and $c^b$, so that $\bm{\gamma}^a=(\zeta^a,\nu^a)$,
$\bm{\gamma}^b=(\zeta^b,\nu^b)$, with 
$\zeta^a>0$ and $\zeta^b>0$ the correlation parameters,
and $\nu^a$ and $\nu^b$ the degrees of freedom. The four series in Figure~\ref{Fig:Persistent_processes}
were simulated using such a copula for $c_2$ with 
$w=0.5, \zeta^a=\zeta^b=0.9,\nu^a=\nu^b=3$. Each element $u_t$ was transformed
to $y_t$ using the quantile functions of the four marginal
distributions.

In our empirical work we use either t-copulas for $c^a$ and $c^b$, 
or `convex Gumbels' defined as follows.
Let $c^G(u,v;\tau)$ be the density of a Gumbel copula parameterized (uniquely)
in terms of its Kendall tau $\tau\geq 0$. 
Then the convex Gumbel has a density $c^{cG}$ equal to the convex
combination of that of the Gumbel and it's rotation 180 degrees (ie. the
survival copula), so that
\[
c^{cG}(u,v;\tau,\delta)=\delta c^G(u,v;\tau) + (1-\delta)c^G(1-u,1-v;\tau)\,,
\]
with $0\leq \delta \leq 1$. This copula was suggested by~\citet{junker2005}.
When employed in Equation~(\ref{eq:cmix}), it gives a
five parameter bivariate copula with $\bm{\gamma}^a=(\delta^a,\tau^a)$, 
$\bm{\gamma}^b=(\delta^b,\tau^b)$,
and a density $c^{MIX}$ that is equal 
to a mixture of all four 90 degree rotations of the Gumbel, similar to the 
jointly symmetric copula
of~\citet{oh2016} in the bivariate case. Table~\ref{tab:copfns}
gives the copula functions $C^{MIX},C^{cG}$
for both the mixture and convex Gumbel copulas.

To show that $C^{MIX}$
can reproduce the features exhibited by the empirical
copulas in Figure~\ref{Fig:Mixture_ofCop_densities}(a--d),
we fit it (with t copula components) to the same four copula datasets. 
The parameters $\bm{\gamma}$ are estimated
by maximizing the copula density at
Equation~(\ref{eq:coplike}), which is the likelihood conditional
on the copula data (the point estimates are reported in the Online Appendix).
Figure~\ref{Fig:Mixture_ofCop_densities}(e--h) plots $c_2$ for the four
estimated copulas, and
in each case the mixture copula reproduces the shape of the corresponding 
empirical copula well.
To show the mixture copulas also replicate
the quantile dependence,
we compute the quantile dependence coefficients
$\lambda_{\mbox{\tiny low}}(\alpha) =P(u_t<\alpha|u_{t-1}<\alpha)$
and $\lambda_{\mbox{\tiny up}}(\alpha)  = P(u_t>\alpha|u_{t-1}>\alpha)$, 
for both the fitted 
mixture copulas and the empirical copulas. 
Figure~\ref{Fig:Tal_dep_arch1} plots these coefficients against $\alpha$, where 
$\lambda_{\mbox{\tiny low}}(\alpha)$ is plotted for $0<\alpha<0.5$, 
and $\lambda_{\mbox{\tiny up}}(\alpha)$ for $0.5<\alpha<1$.
The
coefficients of the mixture and empirical copulas are very close.
Figure~\ref{oafig:offdiag} in the Online Appendix shows that the
same is true
for the quantile dependence coefficients in the off-diagonal quadrants.

For Markov processes of order $p>1$, 
we follow \citet{SmithMinAlmCza2010} 
and use a drawable vine (or `D-vine'). 
A vine copula density is equal to
the product of the densities of a sequence of bivariate copula components, 
called 
`pair-copulas' (\citealp{Aas2009182}).
In a general D-vine there are $T(T-1)/2$ of these, although in our
stationary time series
case there are only $p$ unique pair-copulas and a single ordering
of the variables (i.e. the time order).
To define the D-vine, for $s<t$
denote $u_{t|s} = F(u_t|u_{s},\dots,u_{t-1})$,
 $u_{s|t} = F(u_s|u_{s+1},\dots,u_{t})$
 and $u_{t|t} = u_t= F_t(y_t)$. Then, as shown in Appendix~\ref{append:specdvine},  
the D-vine copula density is
\begin{eqnarray}
c^{DV}(\boldsymbol{u}) &= &\prod_{t=2}^T f(u_t|u_{\max(1,t-p)},\ldots,u_{t-1}) \nonumber \\
&= &\prod_{t=2}^{T}\prod_{k = 1}^{\text{min}(t-1,p)}
c_{k+1}\left(u_{t-k|t-1},u_{t|t-k+1};\bm{\gamma}_{k+1}\right)\,.
\label{eq:DvineLikeli}
\end{eqnarray}
When $p=1$, $c^{DV}$ is equal to the density at
Equation~(\ref{eq:coplike}).
Each pair-copula density 
$c_{k+1}$ has a parameter vector $\bm{\gamma}_{k+1}$, which we
denote explicitly. 
When $k>1$, $c_{k+1}$
captures dependence between $y_{t-k}$ and $y_{t}$,
conditional on the intervening observations
$\left(y_{t-k+1},\dots,y_{t-1}\right)$. When the series is strongly
stationary, the bivariate distribution of 
$y_{t-k},y_{t}|y_{t-k+1},\ldots,y_{t-1}$ does not vary with $t$, so that neither does
the pair-copula density $c_{k+1}$ nor the parameters $\bm{\gamma}_{k+1}$.
Throughout this paper we use
mixture copulas with densities given at Equation~(\ref{eq:cmix}) for
each of
the pair-copula components $c_2,\dots,c_{p+1}$. 
Therefore, each pair-copula $c_{k+1}$
has parameter vector $\bm{\gamma}_{k+1}=\{w_{k+1},\bm{\gamma}_{k+1}^a,\bm{\gamma}_{k+1}^b\}$, and the vine copula density
$c^{DV}$ has
parameters $\bm{\gamma}=\{\bm{\gamma}_2,\ldots,\bm{\gamma}_{p+1}\}$. 
Last, the pair-copula arguments $u_{t|s},u_{s|t}$ are computed from $\bm{u}$
using the efficient algorithm in Appendix~\ref{append:argevalUnivariate}.

We show how the vine copula can replicate the serial dependence
characteristics of three ARCH(3) models in the Online Appendix.

\vspace{-15pt}
\subsection{Measuring Persistence in Volatility}\label{sec:2}
\vspace{-5pt}
We measure serial dependence in the series values using the bivariate marginal
copulas
\[
	\bar{c}(u_{t-k},u_{t})=\int c^{DV}(\bm{u})du_{j\notin \{t-k,t\}}\,,
\] 
for $k\geq 1$. When $k=1$, the marginal copula
is simply the pair-copula
$c_2(u_{t-1},u_{t};\bm{\gamma}_2)$. When $k>1$, the marginal copulas
are unavailable in closed form, but can
be computed via simulation from the D-vine; see
\citet{SmithMinAlmCza2010} for details on how to simulate from a vine copula. 
However, popular
pairwise dependence measures computed from
these marginal copulas do not measure volatility persistence.
For example, for the ARCH and SV processes above,
both Spearman's rho and Kendall's tau of $c_2$ are exactly zero.

We therefore propose new measures of volatility persistence.
These
are computed from the
bivariate copulas
of $(v_{t-k},v_{t})$, for $k\geq 1$, 
where $v_t=V(y_t-E(y_t))$ is a transformation of the mean-corrected time series values.
The smooth transformation
$V:\mathbb{R} \rightarrow \mathbb{R}^+$ is defined so that:
\begin{itemize}
\item[(i)] $V(a) = V(-a) > 0$, and $V(0)= 0$ {\em (symmetry around zero)}, \mbox{ and}
\item[(ii)] $\frac{d}{da}V(a)>0$ if $a>0$, and $\frac{d}{da}V(a)<0$ if $a<0$\,.
\end{itemize}
Examples include $V(a) = |a|$ and $V(a) = a^2$, and we label the copula of 
$(v_{t-k},v_t)$ a `volatility copula'.
Measures of dependence computed from this volatility copula are pairwise
measures of volatility persistence in the time series at lag $k$. 
The copula 
functions and densities of these transformed time series values are
given by the following theorem.

\begin{theorem}\label{Theorem:1}
For $s<t$, let $y_s,y_t$ be time series observations with marginal distribution functions $F_s,F_t$, 
marginal means $\mu_s,\mu_t$, bivariate marginal copula function $\bar C$ and density
 $\bar c$.
Then the copula function of the transformed values $v_s=V(y_s-\mu_s)$, $v_t=V(y_t-\mu_t)$ is
\begin{equation}
C_V(\tilde{u}_s,\tilde{u}_t) =
\sum_{i=1}^{2}\sum_{j=1}^{2}(-1)^i(-1)^j\bar C\left(F_s\left(\mu_s+(-1)^iG({F}^{-1}_{V_s}(\tilde{u}_s))\right),F_t\left(\mu_t+(-1)^jG({F}^{-1}_{V_t}(\tilde{u}_t))\right)\right)\,,
\end{equation}
with corresponding density
\[
c_V(\tilde{u}_s,\tilde{u}_t) = \frac{ \sum_{i=1}^{2}\sum_{j=1}^{2}f\left(\mu_s+(-1)^iG({F}^{-1}_{V_s}(\tilde{u}_s)), \mu_t+(-1)^jG({F}^{-1}_{V_t}(\tilde{u}_t))\right)G'({F}^{-1}_{V_s}(\tilde{u}_s))G'({F}^{-1}_{V_t}(\tilde{u}_t))}{f_{V_s}\left( F_{V_s}^{-1}(\tilde{u}_s)\right)f_{V_t}\left( F_{V_t}^{-1}(\tilde{u}_t)\right)}\,,
\]
\vspace{-10pt}
where
\vspace{-10pt}
\begin{eqnarray*}
F_{V_j}(v_j) &= &F_j(G(v_j)+\mu_j)-F_j(-G(v_j)+\mu_j)\,,\\
f_{V_j}(v_j) &= &\left(f_j(G(v_j)+\mu_j)+f_j(-G(v_j)+\mu_j)\right)G'\left(v_j\right)\,,
\end{eqnarray*}
are the marginal distribution and density functions of $v_j$,
for $j \in \{s,t\}$, $\tilde{u}_j=F_{V_j}(V(y_j-\mu_j))$,
and $G$ is a differentiable function such that $G(V(a)) = |a|$ for any $a \in \mathbb{R}$.\\
{\bf Proof}: See Appendix~\ref{appen:proofcotinouscase}.
\end{theorem}
\vspace{-10pt}
Note that in Theorem~1 above
we do not index $\bar C,C_V,\bar c$ and $c_V$ by $s,t$ to aid readability.

We make a number of observations about the expressions for $C_V$ and $c_V$ in Theorem~1. 
First, they
do not vary with specific choice of 
transformation $V$. Consequently, measures of dependence computed
from this copula are also invariant with respect to $V$, and in this way are
general measures of volatility persistence. 
Second, they can be computed
analytically,
except for the inversion of $F_{V_j}$, which is numerical.
Third, they
apply equally to stationary or non-stationary time series $\{y_t\}$.
However, in the former case, $C_V$ and $c_V$
can be further simplified because the margin is time invariant, so that
$F_s=F_t$ for all $s,t$.
Last, the expressions are
not only a function of
the marginal copula of $(y_s,y_t)$, but
also of the margins $F_s,F_t$. The
implication for applied modeling is that the choice of copula
at Equation~(\ref{eq:copcdf})
does not solely determine the form and degree of persistence in
the volatility of the series $\{y_t\}$.

When both $F_s$ and $F_t$ are symmetric, the expressions for $C_V$ and $c_V$ are
simplified as below.
\begin{lemma}\label{eq:lemma1}
If $F_s$ and $F_t$ are both symmetric, then
\[
C_{V}(\tilde{u}_s,\tilde{u}_{t}) =\sum_{i=1}^{2}\sum_{j=1}^{2}
(-1)^i(-1)^j \bar C\left(\frac{1+(-1)^i\tilde{u}_s}{2},\frac{1+(-1)^j\tilde{u}_{t}}{2}\right)\,,\mbox{ and}
\]
\[
c_{V}(\tilde{u}_{s},\tilde{u}_{t}) =  \frac{1}{4}\sum_{i=1}^{2}\sum_{j=1}^{2}
\bar c\left(\frac{1+(-1)^i\tilde{u}_s}{2},\frac{1+(-1)^j\tilde{u}_{t}}{2}\right)\,.
\]
{\bf Proof}: See Appendix~\ref{appen:lemmasymmetry}
\end{lemma}
\vspace{-10pt}
In Lemma~\ref{eq:lemma1}, the expressions for $C_V$ and $c_V$
do not involve $F_s$ or $F_t$ so that, in this special
case only, the persistence in the volatility of the series is unaffected by the 
choice of marginal distributions.

When $s=t-k$, measures of dependence computed
from $C_V$ are persistence metrics for the volatility
at lag $k\ge 1$. For example, Spearman's rho is
\begin{equation}
\rho_{t-k,t}^v = 12\int\int C_{V}(\tilde{u}_{t-k},\tilde{u}_{t})d\tilde{u_t}d\tilde{u}_{t-k}-3
=12E(\tilde{u}_{t-k}\tilde{u}_{t})-3\,. \label{eq:spear}
\end{equation}
For the D-vine, when $k=1$ the marginal copula for $(y_{t-1},y_{t})$ is
the pair-copula with density $c_2(u_{t-1},u_{t};\bm{\gamma}_2)$.
From this, $C_V$ can be computed
using Theorem~1, and $\rho^v_{t-1,t}$
at Equation~(\ref{eq:spear})
evaluated by bivariate numerical
integration.
However, when $k>1$, the marginal copula for $(y_{t-k},y_{t})$
is unavailable in closed form, and $\rho^v_{t-k,t}$ needs to be evaluated via
Monte Carlo simulation.
We note that because our time series model is stationary, 
it is straightforward to show that 
$\rho^v_{t-k,t}$ does not vary with $t$, so that we simply denote it as
$\rho^v_k$. Last, 
other measures of dependence can
be computed from $C_V$ similarly.

To highlight the coherence of this measure of persistence in volatility, we compute
$\rho^v_1$ for the four heteroskedastic time series used to fit the copulas depicted in
Figure~\ref{Fig:Mixture_ofCop_densities}.
Table~\ref{Tab:SpearmansARCH1SV1} reports these values, along with Spearman's rho between
$(y_{t-1},y_{t})$, 
which we denote as $\rho^y_1$. Both
metrics are computed using numerical integration for
the fitted mixture copulas. For comparison, we also compute
equivalent nonparametric
estimates
of $\rho^v_1$ and $\rho^y_1$ directly from the time series $\{y_t\}$ and $\{v_t\}$.
We make three observations. First,
$\rho^y_1$ is close to zero throughout, and is an inadequate 
measure of serial dependence for these heteroskedastic time series.
Second, $\rho^v_1$ is positive
throughout, and increases as the parameters $\alpha_1$ and $\phi_1$ of
the ARCH(1) and SV(1) models increase. Third, the values for
$\rho^y_1$ and $\rho^v_1$ computed using the fitted parametric mixture copula are
similar to those computed empirically. This is further evidence that the mixture
copula is an adequate parametric model of serial dependence for 
the heteroskedastic 
series. For further comparison, we also report the first order linear
autocorrelations of the series, the absolute values $|y_t|$, and the
squared values $y_t^2$. These are consistent with those from the mixture copula
model, although the autocorrelations of the squared and absolute values differ-- whereas
$\rho_1^v$ is invariant to the form of transformation $V$.
 
\vspace{-15pt}
\subsection{Modeling USD/AUD Exchange Rate}\label{sec:2.4}
\vspace{-5pt}
\subsubsection{First order copula model}
\vspace{-5pt}
To illustrate the advantages of our time series copula model, 
we employ it to model daily returns on 
the USD/AUD exchange rate
from 2 Jan 2001 until 7 Aug 2015,
sourced
from the Federal Reserve Economic Data (FRED) database. 
The series exhibits strong
heteroskedasticity, along with an asymmetric and heavy-tailed
marginal distribution.
Figure~\ref{Fig:AUD_histogram}
plots a histogram of the
$T=3669$ returns, which 
have 
skew of $-0.654$ and kurtosis of $15.15$. 
Also
plotted are the margins of GARCH(1,1), EGARCH(1,1) and GARCH-t(1,1) models
fit to this data, computed by simulation. 
These models are widely used for 
such data~\citep{hansen2005}, yet have margins that are necessarily symmetric 
and inaccurate.
In contrast, we model the margin nonparametrically using the
adaptive
kernel density estimator of \citet{Shimazaki2010}--- also plotted
on Figure~\ref{Fig:AUD_histogram}--- from which the copula data
are computed. The use of a nonparametric time invariant margin, 
combined with a parametric copula, is also advocated
by~\citet{chen2006} and~\citet{chen2009} for stationary Markov series.
We employ the 
first order time series copula at
Equation~(\ref{eq:coplike}), with
the mixture copula for $c_2$, where $c^a,c^b$ are the densities of
bivariate t copulas, so that there are 5 copula parameters.
The resulting copula
model allows for heteroskedastic serial dependence,
but with a margin that 
is consistent with that observed empirically.

We estimate the copula parameters using both MLE
and Bayesian
posterior inference.
For the latter,
flat or uninformative proper priors are used for the copula parameters, 
and computation is by
Markov chain Monte Carlo (MCMC), where the 
parameters were generated as a block using adaptive random walk
Metropolis-Hastings~\citep{roberts2009}.
Table~\ref{Tab:CopulaEstimatesAUD} reports the point estimates for 
both the copula parameters and serial dependence metrics. Also reported
are 90\% confidence intervals for the MLE, along with the 
90\% posterior probability 
intervals. We make the following observations. First, 
while the confidence and posterior intervals
are wide for the parameters, those for
the dependence metrics are not. This is common for copulas with 
multiple parameters, where a wide range of parameter values
can correspond to similar copula functions and dependence
values. Second, the posterior mean and MLE for the dependence metrics are
almost identical. Third,
return values exhibit negligible first order serial
dependence ($\rho_1^y$), but have positive first order extremal
tail dependence
($\lambda_{\mbox{\tiny low}}^y= \lim_{\alpha \to 0} \mbox{Pr}(u_t<\alpha|u_{t-1}<\alpha)$).
Last, our proposed measure of first order volatility persistence ($\rho^v_1$) is positive,
as are the corresponding quantile dependence metrics
($\lambda_{\mbox{\tiny low}}^v(\alpha)= \mbox{Pr}(\tilde u_t<\alpha|\tilde u_{t-1}<\alpha)$
and
$\lambda_{\mbox{\tiny up}}^v(\alpha)= \mbox{Pr}(\tilde u_t>1-\alpha|\tilde u_{t-1}>1-\alpha)$)
computed from the volatility copula in Theorem~1.

Finally, Figure~\ref{fig:AUD_USD_fittedcop} plots the fitted copula
density in panel~(a). For comparison, also plotted in panel~(b)
is the density of
a first order copula
model fitted to the same data, but
where $c^{MIX}$ has convex Gumbel components. Both densities are very similar
and have the `cross shape' that is indicative of serial dependence
in heteroskedastic series.

\vspace{-5pt}
\subsubsection{Validation study}\label{sec:valid}
\vspace{-5pt}
Based on the USD/AUD exchange rate data, 
we undertake a validation study. We fit four time series 
copulas of the form at Equation~(\ref{eq:DvineLikeli}) to the copula data,
as follows:
\begin{itemize}
\item[] Copula A1: An order $p=1$ D-vine with t-copula based mixture components.
\item[] Copula A5: An order $p=5$ D-vine with t-copula based mixture components.
\item[] Copula B1: An order $p=1$ D-vine with convex Gumbel based mixture components.
\item[] Copula B5: An order $p=5$ D-vine with convex Gumbel based mixture components.
\end{itemize}
Copula~A1 is the first order model in Section~\ref{sec:2.4}, to which we add
a higher order D-vine with $p=5$ and component pair-copulas of the same form. 
Copulas~B1 and~B5 are also
D-vines with pair-copula densities given by $c^{MIX}$, each with component
densities $c^a$ and $c^b$ that are
convex Gumbel densities discussed previously. Note that both Copulas~A1 and~B1 are
five parameter copulas, whereas Copulas~A5 and~B5 are parsimonious D-vines with a total 
of $5\times 5=25$ parameters each. The posterior of the copula
models are obtained using MCMC, where the parameters of each 
pair-copula were generated as a block using adaptive
random walk Metropolis-Hastings,
and with blocks generated in random order.
Table~\ref{tab:bench} reports the deviance information criteria (DIC) for each
copula model.
This is computed conditional on the same copula data, 
and is 
$\mbox{DIC}_2$ of~\citet{celeux2006}.
Lower DIC values are preferred, so that longer lag lengths dominate, with Copula~A5 optimal by
this measure.

The ARCH(1), GARCH(1,1), GARCH-t(1,1) and EGARCH(1,1) models, estimated using MLE, are used
as benchmarks. Table~\ref{tab:bench} reports the four
(first order) serial dependence metrics. 
As expected, for all models,
serial dependence in the returns ($\rho_1^y$) is close to zero, and
volatility persistence ($\rho_1^v$) is positive. 
In each model, the first order ($k=1$)
volatility copula exhibits asymmetric and positive 
quantile dependence ($\lambda_{\mbox{\tiny up}}^v(0.05) > \lambda_{\mbox{\tiny low}}^v(0.05) > 0$), 
which is 
something that we repeatedly observe with heteroskedastic series. Interestingly,
the metrics from the copula models are close to those of the GARCH-t(1,1) model,
which is widely considered the most accurate of the benchmark models
for daily exchange rate returns~\citep{baillie2002}.

To judge the quality of the fitted models, we examine
one day ahead VaR coverage as follows. For each fitted 
model, the 1-step ahead
predictive distributions $F_{t|t-1}(y_t)$ are computed 
for days $t=2,\ldots,T$. From these we compute
$VaR_{t|t-1}(\alpha)=F_{t|t-1}^{-1}(\alpha)$, for 
$0<\alpha<1$, along with
the mean
number of exceedences during the $T-1=3668$ days, defined as
$\hat \alpha = \frac{1}{T-1}\sum_{t=2}^T {\rm 1}(y_t<VaR_{t|t-1}(\alpha))$.
Table~\ref{tab:exceed} reports $\hat \alpha$ for 
$\alpha \in \{0.01, 0.05, 0.1, 0.9, 0.95, 0.99\}$, and shows that the copula
models have accurate coverage. 
Figure~\ref{Fig:Back_testing_AUD} plots $\hat \alpha$ from
the Copula~B5 and four GARCH models against $\alpha$, 
for values $0.01<\alpha<0.1$ in panel~(a), and
$0.9<\alpha<0.99$ in panel~(b). 
Deviations from the 
black 45 degree line indicate inaccurate VaR coverage, and it 
can be seen that the
copula model dominates the GARCH models in both tails-- particularly
those with Gaussian innovations. We note that 
the predictive distributions of the GARCH models are necessarily Gaussian or t,
whereas those
from the copula models are not.

Last, in Table~\ref{tab:exceed}
we also report the results of the~\citet{christoff1998} test of correct conditional coverage
for all quantiles and models.
This test 
assesses jointly whether the empirical coverage equates to the corresponding theoretical value and whether
the exceedences are serially independent. 
Models that produce forecasts that fail to reject the conditional coverage test are deemed to perform well 
in predicting VaR. The test results
suggest all four copula models dominate the GARCH benchmarks.

\vspace{-15pt}
\subsection{Simulation Study}\label{sec:simstudy}
\vspace{-5pt}
To illustrate the robustness of the time series copula to model misspecification
we undertake a small simulation study based on the AUD/USD exchange rate data. 
One hundred datasets, each of length
$T=3669$, were simulated from the fitted ARCH(1), Copula~B1, GARCH(1,1) and Copula~B5 models 
in Section~\ref{sec:valid} above.
For each model and dataset we fit both the correct and one
incorrect model (listed in Table~\ref{tab:RRMSE}), 
giving
eight fitted models in total.

To measure the accuracy of the estimated volatility dependence, we 
compute $\rho_1^v,\ldots,\rho_5^v$ from the fitted models. These coefficients
are computed by simulating series of length 1 million from the models, and then computing the sample
Spearman's rho of $V(y_t)$ and $V(y_{t-k})$ for $k=1,\ldots,5$. We repeat this
for all 100 datasets and compute the root mean squared error (RMSE)
for each coefficient and fitted model. Here, the true value of $\rho_k^v$ can be computed accurately
via simulation from the true model as well. Table~\ref{tab:RRMSE} reports the ratio of the RMSE values of the 
misspecified models, relative to that obtained by fitting the correct model. 
Greater relative RMSE values indicate
that the fitted misspecified model does not capture the volatility serial dependence well.
The results indicate that the two copula models reproduce the volatility serial dependence structure
of the GARCH models well, although the converse is not true.
For example, for $\rho^v_1$ the relative RMSEs
of incorrectly fitting the Copula~B1 and~B5 models are
only 1.16 and 1.04; yet the relative RMSEs are 3.41 and 3.25
when incorrectly fitting
the ARCH(1) and GARCH(1,1) models.

%% file: sect3.tex
\vspace{-15pt}
\section{Multivariate Heteroskedastic Time Series}\label{section:3}
\vspace{-10pt}
\subsection{Copula Model}
\vspace{-5pt}
Copulas can also be used to model dependence
in multivariate time series. The copula model for the
$T$ observations
$\bm{y}=(\bm{y}_1',\ldots,\bm{y}_T')'$ of a vector
$\bm{y}_t=(y_{1,t},\ldots,y_{m,t})'$ of $m$ continuous values
has density
\begin{equation}
f(\bm{y})=c(\bm{u})\prod_{t=1}^T\prod_{j=1}^m f_{j,t}(y_{j,t})\,, \label{eq:multipdf}
\end{equation}
where $\bm{u}=(\bm{u}_1',\ldots,\bm{u}_T')'$,
$\bm{u}_t=(u_{1,t},\ldots,u_{m,t})'$,
$u_{j,t}=F_{j,t}(y_{j,t})$, $F_{j,t}$ is the marginal distribution function
of $y_{j,t}$, and $f_{j,t}(y_{j,t})=\frac{d}{d y_{j,t}}F_{j,t}(y_{j,t})$. 
The copula density in Equation~(\ref{eq:multipdf}) is of dimension $mT$,
and captures both serial and cross-sectional dependence
in the series jointly. Selection of an appropriate high-dimensional 
copula is the main
challenge in constructing the model. 

\citet{biller2003}, \citet{biller2009} and~\citet{smithvahey2015} all use
Gaussian copulas, with parameter matrix equal to the
correlation matrix of a stationary
vector autoregression.
However, just as in the univariate case, the Gaussian copula is unable to 
capture the volatility persistence exhibited by heteroskedastic time series.
As an alternative, 
\citet{beare15}, \citet{brechmann15}
and~\citet{Smith2015} all suggest using vine copulas for (strongly) 
stationary
series. We follow these
authors
and employ a D-vine, with
pair-copula components of the form at
Equation~(\ref{eq:cmix}) to capture
heteroskedasticity.
We outline this below, although refer to 
\citet{Smith2015} for further details on the specification of 
the vine and its time series properties.

When the multivariate series is (strongly) stationary and
Markov of lag $p$, the D-vine copula
density can be written as
\begin{equation}
c(\bm{u})=K_0(\bm{u}_1)\prod_{t=2}^T\left( K_0(\bm{u}_t) 
\prod_{k=1}^{\min(t-1,p)} K_k(\bm{u}_{t-k},\ldots,\bm{u}_t)\right)\,.
\label{eq:multicop}
\end{equation}
The functionals $K_0,\ldots,K_p$ are each products of blocks of
pair-copula densities,
and do not vary with $t$ for stationary series. They are
defined as
\[
K_k\left(\boldsymbol{u}_{t-k},\dots,\boldsymbol{u}_t\right) = 
\begin{cases}
\prod_{l_1=1}^{m}\prod_{l_2=1}^{l_1-1}c_{l_2,l_1}^{(0)}\left(u_{j|i-1},u_{i|j+1};\bm{\gamma}_{l_2,l_1}^{(0)}\right)& \text{if} \ \ k=0\\
\prod_{l_1=1}^{m}\prod_{l_2=1}^{m}c_{l_2,l_1}^{(k)}\left(u_{j|i-1},u_{i|j+1};
\bm{\gamma}_{l_2,l_1}^{(k)}\right)& \text{if} \ \ 1\leq k\leq p\,,
\end{cases}
\]
where $c^{(k)}_{l_2,l_1}$ is a bivariate pair-copula density 
with parameters $\bm{\gamma}_{l_2,l_1}^{(k)}$. 
When $k=0$, there are $m(m-1)/2$ of these associated with $K_0$, 
and they collectively capture
cross-sectional dependence between the $m$ variables. For example,
if they were each equal to the bivariate independence copula with 
density $c^{(0)}_{l_2,l_1}=1$, then 
$K_0=1$ and the variables would be independent at
any given point in time. When $k>p$, there
are $m^2$
pair-copulas associated with block $K_k$ that capture serial dependence 
at lag $k$. 
In total,
there are $p(m^2)+m(m-1)/2$ unique pair-copulas, which is much less than
the $Tm(Tm-1)/2$ in an
unconstrained D-vine. 
The indices of the pair-copula 
arguments are $i=l_1+m(t-1)$ and $j=l_2+m(t-k-1)$, 
and
the argument values themselves
$u_{i|j},u_{j|i}$
are computed
using the algorithm in Appendix~\ref{append:argevalMultivariate}.
Last, we note that if $m=1$, then $K_0=1, i=t, j=t-k$ and 
$K_k=c_{1,1}^{(k)}(u_{t-k|t-1},u_{t|t-k+1})$, so that with the notation
$c_{k+1}\equiv c_{1,1}^{(k)}$, the copula densities at 
Equations~(\ref{eq:DvineLikeli}) and~(\ref{eq:multicop}) are the same.
\vspace{-15pt}
\subsection{Estimation, Serial Dependence and Volatility Dependence}
\vspace{-5pt}
Estimation
is similar to 
the univariate case. 
The marginal distribution of each variable is estimated nonparametrically
using adaptive kernel density estimation, from which the copula
data are constructed.
Equation~(\ref{eq:multicop}) gives the likelihood, conditional on the copula data.
It
can be difficult to maximize for higher values of $m$ and $p$, so that
we
follow~\citet{MinCzado2010} and
\citet{SmithMinAlmCza2010} and 
use MCMC to evaluate the posterior distribution.
In the sampling scheme, the parameter
vector of each unique pair-copula is generated jointly,
conditional on the parameters of the 
other pair-copulas. To do so, a Metropolis-Hastings step with an adaptive
multivariate
random walk Gaussian proposal~\citep{roberts2009} is used. Key to implementation
is the efficient computation of the likelihood,
as outlined in Appendix~\ref{append:argevalMultivariate}.

Serial dependence in the series is summarized using measures of 
dependence between pairs $(y_{i,t-k},y_{j,t})$. These
can be arranged into a $(m\times m)$ matrix for 
any given value of $k\geq 0$. Because
the bivariate marginal copula between each pair is
unavailable in closed form, we compute the
metrics via Monte Carlo simulation from the vine copula. This  
can be undertaken efficiently using Algorithm~2 of~\citet{Smith2015}.
For example,
pairwise Spearman's rho can be computed as
\[
\rho^y_{i,j,k}=12E(u_{i,t-k}u_{j,t})-3 \approx \left(\frac{12}{L}\sum_{l=1}^L
u_{i,1}^{[l]}u_{j,k+1}^{[l]}\right) -3\,,
\]
where $\{u_{i,t}^{[l]};\,i=1,\ldots,m,\, t=1,\ldots,k+1\}$ is an iterate 
from the joint distribution of $\{\bm{u}_1,\ldots,\bm{u}_{k+1}\}$,
for $l=1,\ldots,L$. Then the matrix 
$P^y_k=\{\rho_{i,j,k}^y\}_{1\leq i\leq m; 1\leq j\leq m}$ is a
measure of overall
$k$th order serial dependence in the multivariate time series.

Similar dependence measures can be computed
for the pair of volatility proxies $(v_{i,t-k},v_{j,t})$, 
where $v_{i,t}=V(y_{i,t}-\mu_{i,t})$, $\mu_{i,t}=E(y_{i,t})$, and 
$V$ is the function defined in Section~\ref{sec:2}. 
While Theorem~1 is directly applicable here,
the bivariate copula of $(v_{i,t-k},v_{j,t})$ cannot be computed
in closed form because the underlying marginal copula of
$(y_{i,t-k},y_{j,t})$ cannot either. Therefore, we again compute the dependence
measures via Monte Carlo simulation, where iterates of the volatility
proxies can be computed directly from those generated 
for the
series. If 
Spearman's rho of the pair $(v_{i,t-k},v_{j,t})$ is denoted 
as $\rho^v_{i,j,k}$, then these values can be arranged into matrices
$P^v_k=\{\rho_{i,j,k}^v\}_{1\leq i\leq m; 1\leq j\leq m}$,
for a given lag $k\geq 0$. 
The matrix
$P^v_0$ measures cross-sectional dependence in volatility 
at a point in time, with the off-diagonal elements
measuring volatility co-movement. For $k\geq 1$ the leading
diagonal elements of $P^v_k$ are measures of own-series volatility
persistence, whereas the off-diagonals are measures of volatility
spillover. Volatility co-movement and spillover are widely documented
in daily asset and exchange rate returns~\citep{hamao1990,baillie1991},
although multivariate GARCH models usually measure these through conditional
moments, not marginally as we propose here.

\vspace{-15pt}
\subsection{Multivariate Model of Exchange Rates}
\vspace{-5pt}
We extend the analysis of the
USD/AUD exchange rate in Section~\ref{sec:2.4} to include
the 
USD/EUR and USD/JPY rates, which are the two most traded currency
pairs. As before, daily returns
were computed using rates sourced from the FRED database which
are synchronized to New York closing time. The
USD/EUR and USD/JPY returns are both asymmetric
(with skew 0.044 and 0.278) and fat-tailed (with kurtosis 5.23 and
7.22).
Each margin is
modeled nonparametrically using an adaptive kernel density estimator;
see
Figure~\ref{fig:3hist} of the Online Appendix.
The D-vine copula at Equation~(\ref{eq:multicop}) with $p=1$ is used to capture
both cross-sectional and serial dependence simultaneously, with
pair-copula densities given by $c^{MIX}$ at Equation~(\ref{eq:cmix}).
For the components of the mixture copula we use either all t-copulas,
or all convex Gumbels, 
resulting in two D-vines which we label `Copula~A' and `Copula~B', respectively.
Both have a total of $12\times 5=60$
parameters, and their 
posterior mean estimates are reported in 
Table~\ref{Tab:CopulaEstimatesCurrencies}.

To summarize serial dependence in both the series
and its volatility, the top of
Table~\ref{tab:multirhos} reports the posterior mean estimates of the 
matrices $P^y_k$ and $P^v_k$ of pairwise marginal
Spearman's rho.
Results are reported for contemporaneous
($k=0$) and first order serial ($k=1$) dependence.
Results are very similar for both Copulas~A and~B, and 
we make a number of observations. First,
cross-sectional dependence
in the returns ($P^y_0$) is positive throughout, with that between
the USD/AUD and USD/EUR currency pairs being the highest.
Second, there is negligible first order serial dependence 
in the returns
($P^y_1$).
Third, there is co-movement in the volatility 
of the three
series, with positive values on the off-diagonal of $P^v_0$.
Fourth, there is volatility persistence in each series,
with positive values along the leading diagonal of
$P^v_1$. Last, there are positive volatility spillovers between series,
as measured by the off-diagonal elements of $P^v_1$. All five features
are consistent with previous studies of daily exchange rate 
returns~\citep{baillie2002,nakatani2009}.

We compare the dependence matrices with those computed from
two trivariate GARCH models. These are the DCC-GARCH(1,1)~\citep{engle2002}
and BEKK(1,1)~\citep{engle1995} models. The diagonal form
of the BEKK model is used because the likelihood for the full form is not 
log-concave for this series, which is a well-known problem.
Table~\ref{tab:multirhos} also reports
the dependence matrices for these two models, which are computed
via simulation, and
we make four observations. First, all models have positive 
return co-movements $P^y_0$, but are lowest for the DCC-GARCH(1,1).
Second, volatility co-movements $P^v_0$ are also positive
for all models, but are stronger for the copula models.
Third, volatility 
persistence, given by the leading diagonal of $P^v_1$,
is similar in size for all four models. 
Last, the major difference is that the off-diagonals of $P^v_1$ are positive
for the copula models, but almost zero for the two GARCH models. 
Thus, first order volatility spillovers are indicated by the copula
model, but are not by the multivariate GARCH models.

We extend the validation study to include these multivariate models, plus
the BEKK(1,0) model.
We construct the one-day-ahead predictive distributions for the three returns
series and the return on an equally-weighted currency portfolio. 
As in Section~\ref{sec:valid}
we compute the mean number of
exceedances $\hat \alpha$, and plots of these against the quantile $\alpha$
(given in the Online Appendix) suggest
the copula models have more accurate VaR coverage
than the multivariate GARCH models.  For the equally-weighted
currency portfolio, Table~\ref{tab:exceedmulti} reports exceedances for all
five models and the 
results of the Christoffersen tests at six quantiles. As in the univariate case,
the copula models dominate the multivariate GARCH models.

To illustrate the difference in density
forecasts, we plot these for the USD/JPY return on three days
in Figure~\ref{fig:fxdenfore}.
These are the days with the lowest and highest returns
during 
the last 18 months in the data, 
along with the day with return closest to zero.  
Panels~(a--c) plot the densities, while panels (d--f) plot their
logarithm to better visualize the tails. For simplicity,
we only plot the densities
for Copula~A, although for Copula~B are similar.
Note that the multivariate GARCH
models are conditionally Gaussian, 
so that their
predictive distributions are also. In contrast,
the copula forecast densities in panels~(a--c)
are asymmetric, with skew coefficients of 0.113, 0.473 and 0.316, 
respectively.
Panels (d--f) show that the density forecasts
from the copula model also
have heavy tails, with a kurtosis of 5.78, 7.39 and 6.44, respectively.
Clearly, the nonparametric margins combined with the copula function,
translate into non-Gaussian predictive distributions. 

%% file: sect4.tex
\vspace{-15pt}
\section{Discussion}\label{section:5}
\vspace{-5pt}
Time series copula models are very general, in that all time
series models have a copula specification.  For example,~\citet{SmithMan2016}
show how to compute
the copula of a nonlinear state space model numerically.
However, for many 
existing time series models --- including popular models for heteroskedastic
data --- the time series
copula cannot be written in closed form. Our approach
is therefore an alternative copula specification to
capture serial dependence for heteroskedastic data. 
For stationary first order Markov series the bivariate copula density
has mass concentrated along the two diagonals of the unit cube, which mirrors
that found empirically for two popular existing heteroskedastic models.
It is extended 
to higher Markov orders and 
multivariate time series using D-vines. An important observation
is that these vines are 
highly parsimonious, with densities that can be evaluated using
$O(T)$ parallel algorithms.
This enables the copula
models to be readily estimated for the longer series encountered
in practice.

The main theoretical
result is the derivation of the bivariate copula of
a volatility proxy at two time points.
We find that the copula does not 
depend on the specific transformation $V$ used
in Section~\ref{sec:2}.
For example, it is the same
if the volatility proxy is either $v_t=|y_t-E(y_t)|$ or $v_t=(y_t-E(y_t))^2$.
The copula fully characterizes the (unconditional) dependence 
between $v_t$ and $v_{t-k}$ at a given lag $k\geq 1$. While it
is a function of the time series copula, it is also 
a function of the marginal distribution of the data whenever that 
margin is asymmetric. This has an important implication for applied modeling:
while the choice of $C$ in Equation~(\ref{eq:copcdf}) completely
determines the serial dependence of the series, it does not always do so
for volatility.
We show how 
dependence metrics can be computed from the volatility copulas, which provide
measures of volatility persistence, along with co-movement and spillover
for multivariate series. These can be computed
by simulation for any stationary time series, not just the copula
model proposed here. They
can be used to compare the
degree of volatility dependence arising from different nonlinear time series models, 
as in Tables~\ref{tab:bench} and \ref{tab:multirhos}. 

A major advantage of copula models is the simplicity
with which they incorporate complex margins; for example, the
exchange rate returns
series exhibit asymmetry and
heavy tails. As noted by~\citet{chen2006} and others, these can be accurately
captured using nonparametric methods, and we show in
Figure~\ref{fig:fxdenfore} that these
affect the forecast densities substantitally.
In comparison, most existing
time series models are conditionally Gaussian or t distributed, 
and density forecasts
are also; e.g. see \citet{clark2015}. Moreover, the marginal distributions 
are often poorly calibrated, as illustrated in Figure~\ref{Fig:AUD_histogram}.
Ultimately, the VaR forecasts from the copula model are more accurate.
We illustrate this using daily exchange rate returns
with GARCH benchmark models in the univariate case, and 
BEKK and DCC benchmark models in the multivariate case. In the latter,
our copula model also dominates the multivariate GARCH models for
a portfolio of the three rates, indicating that the copula also captures
the cross-sectional dependence accurately.

%% file: append.tex
\newpage
\appendix
\vspace{-15pt}
\section{D-vine Copula Density}
\label{append:specdvine}
\vspace{-10pt}
In this appendix we outline the derivation of the D-vine copula density at Equation~(\ref{eq:DvineLikeli}). 
The copula density of a Markov $p$ process can written as
\[
c^{DV}(\boldsymbol{u}) = \prod_{t=2}^T f(u_t|u_{\max(1,t-p)},\ldots,u_{t-1})\,,
\]
where $f(u_1)=1$ because the marginal distribution of $u_1$ is uniform on $[0,1]$.
For $t-p\leq s < t$, there always exists
a density $c_{t,s}$ on $[0,1]^2$ such that
\begin{eqnarray*}
\lefteqn{f(u_t,u_s|u_{t-1},\ldots,u_{s+1})=f(u_t|u_{t-1},\ldots,u_{s+1})f(u_s|u_{t-1},\ldots,u_{s+1})} & & \\
&\times &c_{t,s}\left(F(u_s|u_{t-1},\ldots,u_{s+1}),F(u_t|u_{t-1},\ldots,u_{s+1});u_{t-1},\ldots,u_{s+1}\right)\,,
\end{eqnarray*}
which is the theorem of Sklar applied conditional on $u_{t-1},\ldots,u_{s+1}$.
In a vine copula, $c_{t,s}$ is a bivariate pair-copula density, and it is
simplified by dropping dependence on $(u_{t-1},\ldots,u_{s+1})$. The pair-copula captures
the dependence between $y_t$ and $y_s$, conditional on the intervening observations. 
Denoting $u_{s|t-1}=F(u_s|u_{t-1},\ldots,u_{s+1})$ and $u_{t|s+1}=F(u_t|u_{t-1},\ldots,u_{s+1})$, the above gives
$f(u_t|u_{t-1},\ldots,u_{s})=c_{t,s}(u_{s|t-1},u_{t|s+1})f(u_t|u_{t-1},\ldots,u_{s+1})$. 
Repeated application
of the above with $s=\max(1,t-p),\ldots,t-1$ gives
\begin{eqnarray*}
f(u_t|u_{\max(1,t-p)},\ldots,u_{t-1}) &= &\prod_{s=\max(1,t-p)}^{t-1}c_{t,s}(u_{s|t-1},u_{t|s+1})\\
 &= &\prod_{k=1}^{\min(p,t-1)}c_{t,t-k}(u_{t-k|t-1},u_{t|t-k+1})\,,
\end{eqnarray*}
where we set $s=t-k$.
If the series is stationary, it is straightforward --- for example, see Smith~(2015) --- to show that 
the pair-copulas $c_{t,t-k}$ are invariant with respect to $t$, so that 
we can write $c_{t,t-k}=c_{k+1}$ throughout, resulting in
Equation~(\ref{eq:DvineLikeli}). 

Last, we note that
compared to Equation~(2.4) of Smith et al.~(2010),
the order of the two arguments of each pair-copula
is switched.
While this
is unimportant when the pair-copula is symmetric, it is when the 
pair-copula is asymmetric, as with the mixture copula $c^{MIX}$ here. It is particularly
important to keep note of the order of the arguments of the pair-copulas when 
implementing the algorithms in Appendix~C.

\vspace{-15pt}
\section{Copula of Transformed Variables}\label{append:coptrans}
\vspace{-10pt}
Consider two continuous
random variables $Y_1$ and $Y_2$, with joint distribution function
$F$, bivariate copula function $\bar C$ and density $\bar c$,
and marginal distribution 
functions $F_1$ and $F_2$, respectively. (In Theorem~1 and Lemma~1, 
these random variables correspond to the time series at times $s$ and $t$,
respectively.)
In this appendix we derive the bivariate
copula function $C_V$ of $V_1=V(Y_1-\mu_1)$ and
$V_2=V(Y_2-\mu_2)$, where $V:\mathbb{R}\rightarrow \mathbb{R}^+$ is the transformation defined
in Section~\ref{sec:2}, $\mu_1=E(Y_1)$ and $\mu_2=E(Y_2)$. We show that,
in general,
$C_V$ is a function of both $\bar C$, and also the marginals $F_1$ and $F_2$.
We
also 
derive the copula density $c_V$ of $V_1$ and $V_2$.
We consider separately the special case where both $Y_1$ and $Y_2$ are
strictly symmetrically distributed.
\vspace{-10pt}
\subsection{General Marginals Case}\label{appen:proofcotinouscase}
\vspace{-5pt}
Let $G(V(a))=|a|$ for any $a\in \mathbb{R}$, and $G(.)$ is a differentiable function
function, where $G:\mathbb{R}^+\rightarrow \mathbb{R}^+$. Recognizing that $G(v_j) = G(V(y_j-\mu_j)) = |y_j-\mu_j|$, the values of $v_j$
can be mapped to $y_j$ (in a one to two mapping), by the identity
$y_j = (-1)^{i} G(v_j)+\mu_j$, for $i \in \{1,2\}$. Since this mapping is deterministic, through the $G$ function, the joint distribution of $(V_1,V_2)$ can be derived from the joint distribution of $(Y_1,Y_2)$:
\begin{eqnarray*}
F_V(v_1,v_2) &= &\mbox{Pr}\left(-G(v_1)< Y_1-\mu_1< G(v_1),
-G(v_2)<Y_2-\mu_2< G(v_2)\right)\\
& = &\sum_{i=1}^{2}\sum_{j=1}^{2}(-1)^i(-1)^jF\left((-1)^iG(v_1)+\mu_1,(-1)^jG(v_2)+\mu_2\right)\,.
\end{eqnarray*}
Further, by Sklar's Theorem, $F(y_1,y_2)=\bar C(F_1(y_1),F_2(y_2))$, so $F_V(v_1,v_2)$ can be written as a function of $\bar C(F_1(y_1),F_2(y_2))$ as
\[F_V(v_1,v_2)=
\sum_{i=1}^{2}\sum_{j=1}^{2}(-1)^i(-1)^j\bar C\left( 
F_1\left((-1)^iG(v_1)+\mu_1\right),
F_2\left((-1)^jG(v_2)+\mu_2\right)\right)\,.
\]
With the marginal distribution function of $V_j$ denoted by $F_{V_j}$ and the corresponding copula datum $\tilde u_j=F_{V_j}(v_j)$, inverting Sklar's theorem yields the copula function 
\begin{eqnarray}
\lefteqn{
C_V(\tilde{u}_1,\tilde{u}_2) = F_V\left(F_{V_1}^{-1}(\tilde{u}_1),
F_{V_2}^{-1}(\tilde{u}_2)\right) } \nonumber\\
 & &=\sum_{i=1}^{2}\sum_{j=1}^{2}(-1)^i(-1)^j\bar C \left(F_1\left((-1)^iG({F}^{-1}_{V_1}(\tilde{u}_1))+\mu_1\right),F_2\left((-1)^jG({F}^{-1}_{V_2}(\tilde{u}_2))+\mu_2\right)\right),
\label{Eq:proof1}
\end{eqnarray}
\vspace{-30pt}
\begin{eqnarray}
\mbox{where }\quad\quad F_{V_j}(v_j)&= &\mbox{Pr}(V_j<v_j)=\mbox{Pr}\left(-G(v_j)<Y_j-\mu_j<G(v_j) \right)\nonumber\\
 &= &F_j\left(G(v_j)+\mu_j\right)-F_j\left(-G_j(v_j)+\mu_j\right). \label{EQ:marginaldist}
\end{eqnarray}
The quantile function $F_{V_j}^{-1}$ can be obtained by numerically 
inverting (\ref{EQ:marginaldist}) for any given marginal $F_j$.

The copula density can be obtained by differentiating the copula function in (\ref{Eq:proof1}):
\begin{eqnarray*}
\lefteqn{c_V(\tilde{u}_1,\tilde{u}_2) = \frac{\partial^2}{\partial\tilde u_1 \partial\tilde u_2}
C_V(\tilde u_1,\tilde u_2)}\\ 
& & =\sum_{i=1}^{2}\sum_{j=1}^{2} \bar c\left(F_1\left(\mu_1+(-1)^iG({F}^{-1}_{V_1}(\tilde{u}_1))\right),F_2\left(\mu_2+(-1)^jG({F}^{-1}_{V_2}(\tilde{u}_2))\right)\right)f_1\left(\mu_1+(-1)^iG({F}^{-1}_{V_1}(\tilde{u}_1))\right)\times   \\ 
& & \ \ \ \ \ \ \ \ \ \ \ \ \ \ \frac{f_2\left(\mu_2+(-1)^jG({F}^{-1}_{V_2}(\tilde{u}_2))\right)G'({F}^{-1}_{V_1}(\tilde{u}_1))G'({F}^{-1}_{V_2}(\tilde{u}_2))}{f_{V_1}\left( F_{V_1}^{-1}(\tilde{u}_1)\right)f_{V_2}\left( F_{V_2}^{-1}(\tilde{u}_2)\right)}\\
& & =\frac{ \sum_{i=1}^{2}\sum_{j=1}^{2}f\left(\mu_1+(-1)^iG({F}^{-1}_{V_1}(\tilde{u}_1)), \mu_2+(-1)^jG({F}^{-1}_{V_2}(\tilde{u}_2))\right)G'({F}^{-1}_{V_1}(\tilde{u}_1))G'({F}^{-1}_{V_2}(\tilde{u}_2))}{f_{V_1}\left( F_{V_1}^{-1}(\tilde{u}_1)\right)f_{V_2}\left( F_{V_2}^{-1}(\tilde{u}_2)\right)}
\end{eqnarray*}
with
\[
f_{V_j}(v_j) =\frac{d}{d v_j}F_{V_j}(v_j) =  \left(f_{j}(G(v_j)+\mu_j)+f_j(-G(v_j)+\mu_j)\right)G'(v_j)\,.
\]
\vspace{-15pt}
\subsection{Symmetric Marginals Case}\label{appen:lemmasymmetry}
\vspace{-5pt}
In the special case where the marginal distributions $F_1$ and $F_2$ are both symmetric around their respective means, we have that $F_j(-G(v_j)+\mu_j)=1-F_j(G(v_j)+\mu_j)$, for $j=\{1,2\}$. Applying this relation to Equation (\ref{EQ:marginaldist}), gives
\begin{equation*}
F_{V_j}(v_j) = 2F_j(G(v_j)+\mu_j)-1.
\end{equation*}
By substituting $F_{V_j}(v_j)=\tilde{u}_j$ and $v_j = F_{V_j}^{-1}(\tilde{u}_j)$, along with simple rearrangements,  
\begin{equation*}
G({F}^{-1}_{V_j}(\tilde{u}_j))=F_j^{-1}\left(\frac{1+\tilde{u}_j}{2}\right)-\mu_j.
\end{equation*}
Since the marginal distribution is symmetric around $\mu_j$, we also have that
\begin{equation*}
-G({F}^{-1}_{V_j}(\tilde{u}_j))=F_j^{-1}\left(\frac{1-\tilde{u}_j }{2}\right)-\mu_j.\\
\end{equation*}
Substituting the simplified expressions for $G({F}^{-1}_{V_j}(\tilde{u}_j))$ and $-G({F}^{-1}_{V_j}(\tilde{u}_j))$ into Equation~(\ref{Eq:proof1}) gives
$$C_V(\tilde{u}_1,\tilde{u}_2) =\sum_{i=1}^{2}\sum_{j=1}^{2}(-1)^i(-1)^j
\bar C\left(\frac{1+(-1)^{i}\tilde{u}_1 }{2},\frac{1+(-1)^{j}\tilde{u}_2 }{2}\right)\,.$$
Finally, by differentiating the copula distribution above, the copula density is 
$$c_V(\tilde{u}_1,\tilde{u}_2) =\sum_{i=1}^{2}\sum_{j=1}^{2}\frac{1}{4}\bar c
\left(\frac{1+(-1)^{i}\tilde{u}_1 }{2},\frac{1+(-1)^{j}\tilde{u}_2 }{2}\right)\,.$$

Note that the copula function of the transformed variable in this special case, where both margins are symmetric around $\mu_j$, does not depend on the form of the marginal distribution $F_j$.

\vspace{-15pt}
\section{Efficient Likelihood Evaluation}\label{sec:appendB}
\vspace{-10pt}
Computing the two D-vine copula densities at
Equations~(\ref{eq:DvineLikeli}) and~(\ref{eq:multicop}) 
requires efficient evaluation of the arguments of the pair-copulas.
In this appendix we outline algorithms to compute these.
The algorithms
are extensions of that orginally proposed
by~\citet{Aas2009182}, and further developed in~\citet{SmithMinAlmCza2010}
and~\citet{Smith2015}.
They differ in three ways:
(i)~they are re-ordered so that the computations can be undertaken in parallel;
(ii)~they exploit the 
parsimonious
structures of the two vine copulas; and~(iii) they are based on recursions 
that account for the pair-copulas being mixtures of
possibly asymmetric copulas.
\vspace{-15pt}
\subsection{Univariate Series}\label{append:argevalUnivariate}
\vspace{-10pt}
The arguments of the pair-copulas
can be computed by exploiting the
recursive relationships
\[
u_{t|s} = h_{s,t}^1(u_{t|s+1}|u_{s|t-1})\,,\mbox{ and } u_{s|t} = h^2_{s,t}(u_{s|t-1}|u_{t|s+1})\,,
\]
where, for the specific vine in Equation~(\ref{eq:DvineLikeli}), if $k=t-s$ then
\[
h^1_{s,t}(v|u) = \frac{\partial}{\partial u} C_{k+1}(u,v;\bm{\gamma}_{k+1})\,,\;\mbox{ and }
h^2_{s,t}(u|v) = \frac{\partial}{\partial v} C_{k+1}(u,v;\bm{\gamma}_{k+1})\,.
\]
Here, $C_{k+1}(u,v;\bm{\gamma}_{k+1})=
\int_0^u \int_0^v c_{k+1}(\tilde u,\tilde v;\bm{\gamma})d\tilde u d\tilde v$
is the pair-copula function for $k=1,2,\ldots,p$.
We note that these recursions are more general than those 
given in~\citet{SmithMinAlmCza2010}. These authors assume that $h_{s,t}^1=h_{s,t}^2$, 
which is true for the pair-copula types they examine. However,
this is not the case when the pair-copula is the mixture copula
with function
$C^{MIX}$, with the partial derivatives given in Table~\ref{tab:copfns}. 

The $O(T^2)$ algorithms in \citet{Aas2009182} and~\citet{SmithMinAlmCza2010} 
compute and store all $T(T-1)$ values 
$\{u_{t|s},u_{s|t}; 1\leq t\leq T, s<t\}$, which is impractical for high
values of $T$. However, to compute the likelihood
in Equation~(\ref{eq:DvineLikeli}), 
only the values ${\cal U}=\{u_{t|s},u_{s|t}; 1\leq t\leq T,\max(1,t-p)\leq s<t\}$ need 
computing and storing. Moreover, we evaluate the elements of ${\cal U}$ in 
a different order
to allow the computations 
to be undertaken in parallel, as follows:

\noindent {\bf Algorithm~1.}\\
\noindent For $t=1,\ldots,T$:\\
\indent Step~(1). Set $u_{t|t}=u_t$.\\
\noindent For $k=1,\ldots,p$:\\
\indent 
For $t=k+1,\ldots,T$ (compute inner loop in parallel):\\
\indent \indent Step~(2.1). $u_{t|t-k}=h^1_{t-k,t}(u_{t|t-k+1}|u_{t-k|t-1})$\\
\indent \indent Step~(2.2). $u_{t-k|t}=h^2_{t-k,t}(u_{t-k|t-1}|u_{t|t-k+1})$\\

Once computed, the elements in ${\cal U}$ need to be stored efficiently.
It is possible to store these in a $(T \times T)$ matrix, with $u_{t|s}$ stored in 
element $(t,s)$, and $u_{s|t}$ in element $(s,t)$. However, this 
is prohibitive for longer time series.
Instead,
${\cal U}$ can be stored efficiently either as a banded matrix with bandwidth $p$,
or a $(T \times p \times 2)$ array, with $u_{t|t-k}$ stored as element $(t,k,1)$, and $u_{t-k|t}$ 
as element $(t,k,2)$. We use the latter approach in our code.

%
\vspace{-10pt}
\subsection{Multivariate Series}\label{append:argevalMultivariate}
\vspace{-5pt}
For the vine copula at Equation~(\ref{eq:multicop}), there is a one-to-one
relationship between the indices $(s,t,l_1,l_2)$, and those of the pair-copula
arguments $(i,j)$.
To evaluate these arguments we use the recursive relationships
\[
u_{i|j} = h_{j,i}^1(u_{i|j+1}|u_{j|i-1})\,,\mbox{ and } u_{j|i} = h^2_{j,i}(u_{j|i-1}|u_{i|j+1})\,.
\]
The functions are
\[
h^1_{j,i}(v|u) = \frac{\partial}{\partial u} C_{l_2,l_1}^{(k)}(u,v;\bm{\gamma}_{l_2,l_1}^{(k)})\,,\;\mbox{ and }
h^2_{j,i}(u|v) = \frac{\partial}{\partial v} C_{l_2,l_1}^{(k)}(u,v;\bm{\gamma}_{l_2,l_1}^{(k)})\,,
\]
where
$s = \ceil{j/m}$, $t = \ceil{i/m}$, $k=t-s$, $l_1 = i-m(t-1)$, $l_2 = j-m(s-1)$, and the pair-copula function
$C_{l_2,l_1}^{(k)}(u,v;\bm{\gamma}_{l_2,l_1}^{(k)})=
\int_0^u \int_0^v c_{l_2,l_1}^{(k)}(\tilde u,\tilde v;\bm{\gamma}_{l_2,l_1}^{(k)})d\tilde u d\tilde v$. As in the 
univariate case, we employ $C^{MIX}$ for the pair-copula functions, so that
the partial derivatives required to compute $h^1_{i,j}$ and $h^2_{j,i}$ above
are given in Table~\ref{tab:copfns}.
Following~\citep{Smith2015}, we 
note that only the values
\[
{\cal U}=\{u_{i|j},u_{j|i}; 1\leq i \leq Tm,\max\allowbreak \left(1,\allowbreak m\left(\ceil{i/m}-1\allowbreak-p\right)+1\right)\leq j<i\}
\]
are needed to compute the likelihood. These can be computed using the
$O(pm^2T)$ algorithm below. 

\noindent {\bf Algorithm~2.}

\noindent For $t=1,\ldots,T$, $l=1,\ldots,m$:\\
\indent 
Step~(1.1). Set $i=t+(l-1)m$.\\
\indent
Step~(1.2). Set $u_{i|i}=u_{l,t}$.\\
\noindent For $r = 1,\dots,(p+1)m-1$:\\
\indent 
For $i = r+1,\dots,mT$ (compute inner loop in parallel):\\
\indent \indent Step~(2.1).
Set $j = i- r$, $s = \ceil{j/m}$, $t = \ceil{i/m}$, $k = t- s$, $l_1 = i-m(t-1)$, $l_2 = j-m(s-1)$.\\
\indent \indent Step~(2.2). Compute $u_{i|j} = h_{j,i}^1(u_{i|j+1}|u_{j|i-1})$.\\
\indent \indent Step~(2.3). Compute $u_{j|i} = h^2_{j,i}(u_{j|i-1}|u_{i|j+1})$.\\

The arguments in ${\cal U}$ are efficiently stored in a 3-dimensional $\left(Tm \times \left(m(p+1)-1\right) \times 2\right)$ array,
with $u_{j|i}$ stored as element $(i,i-j,1)$, and $u_{j|i}$ 
as element $(i,i-j,2)$.

%% file: tabs.tex
\begin{table}[h]
	\setlength\extrarowheight{5pt}
	\centering
	\begin{tabular}{l} \hline \hline
		\underline{Bivariate Mixture Copula (Parameters $\bm{\gamma}=\{w,\bm{\gamma}^a,\bm{\gamma}^b\}$)} \\
		{\em (i)~Copula Function} \\
		$C^{MIX}(u,v;\bm{\gamma})= w C^a(u,v;\bm{\gamma}^a)+(1-w)(v-C^b(1-u,v;
		\bm{\gamma}^b))$ \\
		{\em (ii)~Partial Derivatives} \\
		$h^{MIX,1}(v|u;\bm{\gamma})\equiv \frac{\partial}{\partial u} C^{MIX}(u,v;\bm{\gamma}) = w h^{a,1}(v|u;\bm{\gamma}^a)+(1-w)h^{b,1}(v|1-u;
		\bm{\gamma}^b)$ \\
		$h^{MIX,2}(u|v;\bm{\gamma})\equiv \frac{\partial}{\partial v} C^{MIX}(u,v;\bm{\gamma}) = w h^{a,2}(u|v;\bm{\gamma}^a)+(1-w)(1-h^{b,2}(1-u|v;
		\bm{\gamma}^b))$ \\
		where $h^{x,2}(u|v;\bm{\gamma}^x)\equiv \frac{\partial}{\partial v}C^x(u,v;\bm{\gamma}^x)$ and 
		$h^{x,1}(v|u;\bm{\gamma}^x)\equiv \frac{\partial}{\partial u}C^x(u,v;\bm{\gamma}^x)$ for $x=a,b$\\ \hline
		\underline{Convex Gumbel Copula (Parameters $0\leq \delta \leq 1$, $\tau\geq 0$)} \\ 
		{\em (i)~Copula Function} \\
		$C^{cG}(u,v;\tau,\delta) = \delta C^G(u,v;\tau)+(1-\delta)(u+v-1+C^G(1-u,1-v;\tau))$\\
		{\em (ii)~Partial Derivatives} \\
		$h^{cG}(v|u;\tau,\delta)\equiv \frac{\partial}{\partial u} C^{cG}(u,v;\tau,\delta) = \delta h^G(v|u;\tau)+(1-\delta)(1-h^G(1-v|1-u;\tau))$ \\
		$h^{cG}(u|v;\tau,\delta)\equiv \frac{\partial}{\partial v} C^{cG}(u,v;\tau,\delta) = \delta h^G(u|v;\tau)+(1-\delta)(1-h^G(1-u|1-v;\tau))$ \\
		where $h^G(u|v;\tau) \equiv \frac{\partial}{\partial v} C^G(u,v;\tau) = \frac{\partial}{\partial v}C^G(v,u;\tau)$\\
		\hline\hline
	\end{tabular}
	\caption{Distribution functions and their partial derivatives for the
		mixture copula with density $c^{MIX}(u,v;\bm{\gamma})$, and the
		convex Gumbel copula with density
		$c^{cG}(u,v;\tau,\delta)$. Here, $C^G$ is the Gumbel copula function parameterized
		in terms of Kendall's tau $\tau$. We note that when $C^a$ or $C^b$ are t-copulas, their
		partial derivatives can be found in Aas et al.~(2009).}
	\label{tab:copfns}
\end{table}

\begin{table}[H]
	\centering
	\begin{tabular}{llrrrrrrr}
		\toprule
		     &                    &         \multicolumn{2}{c}{$\rho^y_1$}          &         \multicolumn{2}{c}{$\rho^v_1$}          &                $r_1^y$ &           $ r_1^{|y|}$ &            $r_1^{y^2}$ \\ \midrule
		     & Case               &                Mixture &              Empirical &                Mixture &              Empirical &                        &                        &  \\ \midrule
		ARCH & $(\alpha_1 = 0.5)$ &                 -0.002 &                 -0.003 &                  0.241 &                  0.240 &                 -0.007 &                  0.373 &                  0.451 \\
		     &                    & \footnotesize{(0.022)} & \footnotesize{(0.005)} & \footnotesize{(0.075)} & \footnotesize{(0.005)} & \footnotesize{(0.005)} & \footnotesize{(0.005)} & \footnotesize{(0.005)} \\
		     & $(\alpha_1 = 0.9)$ &                 -0.001 &                 -0.002 &                  0.393 &                  0.371 &                 -0.014 &                  0.655 &                  0.574 \\
		     &                    & \footnotesize{(0.013)} & \footnotesize{(0.005)} & \footnotesize{(0.115)} & \footnotesize{(0.006)} & \footnotesize{(0.005)} & \footnotesize{(0.005)} & \footnotesize{(0.005)} \\ \midrule
		SV   & $(\phi_1 = 0.5)$   &                 -0.004 &                 -0.004 &                  0.186 &                  0.230 &                 -0.005 &                  0.221 &                  0.102 \\
		     &                    & \footnotesize{(0.025)} & \footnotesize{(0.005)} & \footnotesize{(0.018)} & \footnotesize{(0.005)} & \footnotesize{(0.005)} & \footnotesize{(0.005)} & \footnotesize{(0.005)} \\
		     & $(\phi_1 = 0.9)$   &                 -0.002 &                 -0.004 &                  0.395 &                  0.453 &                 -0.026 &                  0.397 &                  0.142 \\
		     &                    & \footnotesize{(0.293)} & \footnotesize{(0.005)} & \footnotesize{(0.206)} & \footnotesize{(0.005)} & \footnotesize{(0.005)} & \footnotesize{(0.005)} & \footnotesize{(0.005)} \\ \bottomrule
	\end{tabular}
	\caption{Spearman's rho of first order serial dependence in the level ($\rho^y_1$) and 
		volatility ($\rho^v_1$) for four datasets simulted from ARCH(1) and SV(1) models. 
		Columns labelled `Mixture' show values for the fitted parametric model, where $c_2$ is modelled with
		the mixture copula.
		Columns labelled `Empirical'
		show nonparametric empirical values. Standard errors are 
	given below in parentheses. For comparison, the
	final three columns report the first order sample autocorrelations
	for the three series $\{y_t\}$, $\{|y_t|\}$ and $\{y_t^2\}$.}
	\label{Tab:SpearmansARCH1SV1}
\end{table}

\begin{table}[H]
	\centering
	\begin{tabular}{lccccc}
		\toprule
		Parameter          &                         $\zeta^a>0$                          &                  $\nu^a$                  &               $\zeta^b>0$               &                 $\nu^b$                 &                   $w$                   \\ \midrule
		Posterior Mean     &                            0.153                             &                   9.668                   &                  0.170                  &                  9.866                  &                  0.474                  \\
		Posterior Interval &           \textit{{\footnotesize	(0.008,0.463) }}            & \textit{{\footnotesize	 (3.638,20.210) }} & \textit{{\footnotesize	 (0.016,0.494)}} & \textit{{\footnotesize	(4.111,21.517)}} & \textit{{\footnotesize	(0.044,0.932)} } \\
		MLE                &                              0                               &                  39.995                   &                  0.020                  &                  4.777                  &                  0.321                  \\
		SE                 &                    \footnotesize{(0.339)}                    &          \footnotesize{(0.171)}           &         \footnotesize{(0.172)}          &         \footnotesize{(0.502)}          &        \footnotesize{( 0.164) }         \\
		Conf. Interval     &              \textit{{\footnotesize(0,0.559)}}               &  \textit{{\footnotesize(39.721,40.269)}}  &    \textit{{\footnotesize(0,0.295)}}    & \textit{{\footnotesize (3.974,5.580)}}  &  \textit{{\footnotesize(0.059,0.583)}}  \\ \midrule
		Metric             & $\lambda^y_{\mbox{\tiny low}} =\lambda^y_{\mbox{\tiny up}} $ &                $\rho^y_1$                 &               $\rho^v_1$                &  $\lambda_{\mbox{\tiny low}}^v(0.05) $  &  $\lambda_{\mbox{\tiny up}}^v(0.05) $   \\ \midrule
		Posterior Mean     &                            0.030                             &                  -0.012                   &                  0.090                  &                  0.054                  &                  0.142                  \\
		Posterior Interval &            \textit{{\footnotesize	(0.012,0.051)}}            &  \textit{{\footnotesize	(-0.041,0.016)}}  & \textit{{\footnotesize	(0.071,0.109)} } & \textit{{\footnotesize	(0.053,0.055)}}  & \textit{{\footnotesize	(0.120,0.164)} } \\
		MLE                &                            0.0395                            &                  -0.012                   &                  0.090                  &                  0.054                  &                  0.147                  \\
		SE                 &                   \footnotesize{ (0.026) }                   &          \footnotesize{(0.020)}           &         \footnotesize{ (0.016)}         &        \footnotesize{ (0.0007) }        &         \footnotesize{ (0.017)}         \\
		Conf. Interval     &              \textit{{\footnotesize(0,0.082)}}               &  \textit{{\footnotesize(-0.045 ,0.021)}}  & \textit{ {\footnotesize(0.064,0.116)}}  & \textit{{\footnotesize(0.053,0.055) }}  & \textit{ {\footnotesize(0.119,0.175)}}  \\ \bottomrule
	\end{tabular}
	\caption{Estimates of the mixture copula parameters (upper half), and corresponding first order serial dependence metrics (lower half),
for the USD/AUD exchange rate series. Both
Bayesian posterior mean and MLEs are reported. Also reported for the former are 90\% probability intervals, and for the latter, standard
errors (SE) and asymptotic 90\% confidence intervals constrained to feasible regions.
The metrics include Spearman's rho for  dependence in the series ($\rho^y_1$) and volatility ($\rho^v_1$), extremal tail dependence
in the series
($\lambda^y_{\mbox{\tiny low}}=\lambda^y_{\mbox{\tiny up}}$), and
lower and upper quantile dependence in the volatility at quantile $\alpha=0.05$,
($\lambda_{\mbox{\tiny low}}^v(0.05) $ and $\lambda_{\mbox{\tiny up}}^v(0.05)$).}
\label{Tab:CopulaEstimatesAUD}
\end{table}     

\begin{table}[H]
	\centering
	\begin{tabular}{lccccc}
		\toprule
		\multirow{1}{*}{Model} & \multirow{1}{*}{$\rho^y_1$} & \multirow{1}{*}{$\rho^v_1$} & \multirow{1}{*}{$\lambda_{\mbox{\tiny low}}^v(0.05) $ } & \multirow{1}{*}{$\lambda_{\mbox{\tiny up}}^v(0.05) $} & \multirow{1}{*}{$\mbox{DIC}_2$} \\ \midrule
		Copula A1              &           -0.012            &            0.090            &                          0.054                          &                         0.147                         &             -68.65              \\
		Copula A5              &           -0.001            &            0.074            &                          0.054                          &                         0.127                         &             -286.42             \\
		Copula B1              &           -0.012            &            0.079            &                          0.054                          &                         0.148                         &             -67.36              \\
		Copula B5              &            0.003            &            0.066            &                          0.053                          &                         0.127                         &             -277.48             \\
		ARCH(1)                &           -0.002            &            0.115            &                          0.055                          &                         0.174                         & --- \\
		GARCH(1,1)             &              0              &            0.137            &                          0.057                          &                         0.173                         & --- \\
		EGARCH(1,1)            &           -0.002            &            0.115            &                          0.056                          &                         0.174                         & --- \\
		GARCH-t(1,1)           &           -0.001            &            0.084            &                          0.055                          &                         0.138                         & --- \\ \bottomrule
	\end{tabular}
	\caption{Dependence metrics for copula and GARCH models fit to the USD/AUD exchange rate returns. These were
computed using numerical integration for the first order Copula~A1 and~B1 models. For all other models, the metrics
were computed by simulation. The Deviance Information Criterion ($\mbox{DIC}_2$) is reported for the copula
models, but not the GARCH models because they are estimated by MLE.}
\label{tab:bench} 
\end{table}     

\begin{table}[H]
	\centering
	\begin{tabular}{lcccccc}
		\toprule
		                       &                                     \multicolumn{6}{c}{Quantile $\alpha$}                                     \\              
		\multirow{1}{*}{Model} &       1\%       &       5\%       &       10\%       &       90\%       &       95\%       &       99\%       \\ \midrule             
		Copula A1              &     0.93\%      &     4.77\%      &      9.27\%      &     90.59\%      &     95.50\%      &     99.15\%      \\              
		Copula A5              &     1.04\%      &     5.18\%      &      9.73\%      & 90.05\% & 95.01\% &     99.15\%      \\              
		Copula B1              &     0.95\%      & 4.93\% &      9.62\%      &     90.29\%      &     95.39\%      &     99.13\%      \\              
		Copula B5              & 1.01\% & 5.07\% &      9.65\%      &     90.21\%      &     95.23\%      &     99.24\%      \\              
		ARCH(1)                &    1.66\%**     &     4.36\%      &     7.28\%**     &    92.48\%**     &    96.37\%**     & 99.05\% \\              
		GARCH(1,1)             &    1.61\%**     & 5.07\% &     9.11\%*      &    91.63\%**     &     95.86\%*     &     99.18\%      \\              
		EGARCH(1,1)            &    1.69\%**     &     4.88\%      &      9.43\%      &    91.38\%**     &     95.64\%*     &     99.18\%      \\              
		GARCH-t(1,1)           &     1.28\%      &     5.62\%      & 10.22\% &     90.13\%*     &     95.39\%      &     99.37\%*     \\ \bottomrule             
	\end{tabular}
	\caption{Mean exceedances $\hat \alpha$ (in percent)
		over $T-1=3668$ days of 
		one day ahead VaR forecasts of USD/AUD exchange rate returns. Results
		are given for eight models and six different quantile values.
		Rejection of the null hypothesis of the conditional coverage Christoffersen~(1998) test is 
		denoted with `*' and `**' at the 95\% and 99\% level of 
		confidence, respectively.}
	\label{tab:exceed}
\end{table}

\begin{table}[H]
	\centering
	\begin{tabular}{ccccc}
		\toprule
		& \multicolumn{4}{c}{Fitted Model/Correct Model} \\ \cline{2-5} 
		Volatility serial & Copula B1/ &  ARCH(1)/  & Copula B5/  & GARCH(1,1)/ \\ 
		dependence  &  ARCH(1)  & Copula B1 & GARCH(1,1) & Copula B5 \\ \midrule
		$\rho_1^v$    &   1.16   &   3.41   &   1.04    &   3.25    \\
		$\rho_2^v$    &   0.99   &   4.77   &   1.10    &   1.22    \\
		$\rho_3^v$    &   0.89   &   6.64   &   1.07    &   1.64    \\
		$\rho_4^v$    &   0.79   &   5.52   &   1.08    &   2.36    \\
		$\rho_5^v$    &   0.98   &   2.06   &   1.06    &   2.32    \\ \bottomrule
	\end{tabular}
	\caption{Relative RMSE of the estimates of $\rho^v_1,\ldots,\rho^v_5$ of the fitted model over the correct model. 
	Four mis-specified models are considered.}
	\label{tab:RRMSE}
\end{table}

\begin{landscape}    
\begin{table}[H]
	\centering
	{\small
	\begin{tabular}{cccccccccccccc}
		\toprule
		                          &                 \multicolumn{6}{c}{Copula A: $C^{MIX}$ with t-Copula Components}                 &  &                \multicolumn{6}{c}{Copula B: $C^{MIX}$ with Convex Gumbel Components}                 \\
		       Parameters         & $\zeta^a>0$ & $\nu^a$ & $\zeta^b>0$ & $\nu^b$ &  $w$  &            Spearman            &  & $\tau^a>0$ & $\delta^a$ & $\tau^b>0$ & $\delta^b$ &  $w$  &            Spearman            \\ \midrule
		$\bm{\gamma}_{1,2}^{(0)}$ &    0.707    & 29.427  &    0.132    &  5.992  & 0.419 &             0.219              &  &   0.457    &   0.576    &   0.164    &   0.280    & 0.518 &             0.211              \\
		                          &             &         &             &         &       &  {\footnotesize(0.197,0.242)}  &  &            &            &            &            &       & {\footnotesize( 0.190, 0.233)} \\
		$\bm{\gamma}_{1,3}^{(0)}$ &    0.504    & 27.729  &    0.133    & 30.852  & 0.646 &             0.271              &  &   0.238    &   0.625    &   0.227    &   0.709    & 0.867 &             0.266              \\
		                          &             &         &             &         &       &  {\footnotesize(0.250,0.292)}  &  &            &            &            &            &       & {\footnotesize( 0.246, 0.285)} \\
		$\bm{\gamma}_{2,3}^{(0)}$ &    0.633    &  7.285  &    0.623    & 30.261  & 0.972 &             0.576              &  &   0.439    &   0.527    &   0.400    &   0.766    & 0.966 &             0.572              \\
		                          &             &         &             &         &       &  {\footnotesize(0.561,0.591)}  &  &            &            &            &            &       & {\footnotesize( 0.558, 0.586)} \\ \midrule
		$\bm{\gamma}_{1,1}^{(1)}$ &    0.082    & 22.012  &    0.297    & 16.473  & 0.642 &             -0.012             &  &   0.025    &   0.347    &   0.357    &   0.702    & 0.834 &             -0.017             \\
		                          &             &         &             &         &       & {\footnotesize(-0.031,0.007)}  &  &            &            &            &            &       & {\footnotesize(-0.038, 0.004)} \\
		$\bm{\gamma}_{1,2}^{(1)}$ &   0.4122    & 11.950  &    0.076    & 25.062  & 0.161 &             -0.030             &  &   0.481    &   0.463    &   0.039    &   0.304    & 0.101 &             -0.023             \\
		                          &             &         &             &         &       & {\footnotesize(-0.050,-0.007)} &  &            &            &            &            &       & {\footnotesize(-0.044,-0.002)} \\
		$\bm{\gamma}_{1,3}^{(1)}$ &    0.070    & 23.628  &    0.130    & 19.198  & 0.678 &             0.007              &  &   0.078    &   0.419    &   0.053    &   0.532    & 0.458 &             0.003              \\
		                          &             &         &             &         &       & {\footnotesize(-0.011,0.027)}  &  &            &            &            &            &       & {\footnotesize(-0.014, 0.021)} \\
		$\bm{\gamma}_{2,1}^{(1)}$ &    0.288    & 14.012  &    0.087    & 22.017  & 0.362 &             0.009              &  &   0.207    &   0.303    &   0.106    &   0.334    & 0.451 &             0.014              \\
		                          &             &         &             &         &       & {\footnotesize(-0.010,0.030)}  &  &            &            &            &            &       & {\footnotesize(-0.006, 0.034)} \\
		$\bm{\gamma}_{2,2}^{(1)}$ &    0.157    & 13.245  &    0.163    & 10.867  & 0.471 &             -0.005             &  &   0.155    &   0.583    &   0.127    &   0.551    & 0.438 &             -0.007             \\
		                          &             &         &             &         &       & {\footnotesize(-0.027,0.015)}  &  &            &            &            &            &       & {\footnotesize(-0.029, 0.014)} \\
		$\bm{\gamma}_{2,3}^{(1)}$ &    0.076    & 20.268  &    0.440    & 15.732  & 0.674 &             -0.031             &  &   0.042    &   0.355    &   0.373    &   0.670    & 0.779 &             -0.033             \\
		                          &             &         &             &         &       & {\footnotesize(-0.054,-0.009)} &  &            &            &            &            &       & {\footnotesize(-0.054,-0.010)} \\
		$\bm{\gamma}_{3,1}^{(1)}$ &    0.211    & 19.605  &    0.252    & 11.745  & 0.555 &             -0.013             &  &   0.050    &   0.552    &   0.290    &   0.607    & 0.759 &             -0.013             \\
		                          &             &         &             &         &       & {\footnotesize(-0.033,0.007)}  &  &            &            &            &            &       & {\footnotesize(-0.034, 0.008)} \\
		$\bm{\gamma}_{3,2}^{(1)}$ &    0.275    & 14.398  &    0.125    & 16.701  & 0.329 &             -0.020             &  &   0.144    &   0.070    &   0.113    &   0.425    & 0.416 &             -0.014             \\
		                          &             &         &             &         &       & {\footnotesize(-0.042,0.001)}  &  &            &            &            &            &       & {\footnotesize(-0.036, 0.007)} \\
		$\bm{\gamma}_{3,3}^{(1)}$ &    0.178    & 18.623  &    0.100    & 25.462  & 0.396 &             -0.004             &  &   0.092    &   0.303    &   0.168    &   0.388    & 0.555 &             -0.005             \\
		                          &             &         &             &         &       & {\footnotesize(-0.025,0.015)}  &  &            &            &            &            &       & {\footnotesize(-0.026, 0.016)} \\ \bottomrule
	\end{tabular}
}
	\caption{Posterior means of the pair-copula parameters
for the D-vines fit to the three-dimensional exchange rate return series. The lefthand side gives the pair-copula
parameters for Copula A, and the righthand side for Copula B.
The
posterior mean and 90\% probability intervals
are also given for the
Spearman's rho of each pair copula.
The USD/JPY, USD/AUD and USD/EUR
returns are denoted
as series 1, 2 and 3, respectively.}
	\label{Tab:CopulaEstimatesCurrencies}
\end{table}
\end{landscape}

\begin{landscape}
\begin{table}[H]
	\centering
	\begin{tabular}{lrrrrrrrrrrrrrrr}
		\toprule
		                                                \multicolumn{8}{c}{Copula Time Series Model with Copula A}                                                  &                          \multicolumn{8}{c}{Copula Time Series Model with Copula B}                           \\ \cline{2-8}\cline{10-16}
		                                   & $JPY_t$ & $AUD_t$ &                           $EUR_t$ &  & $JPY_t$ & $AUD_t$ &                                $EUR_t$ &  & $JPY_t$ & $AUD_t$ &                           $EUR_t$ &  & $JPY_t$ & $AUD_t$ &                                $EUR_t$ \\ \cline{2-4}\cline{6-8}\cline{10-12}\cline{14-16}
		                                   & \multicolumn{3}{c}{$P^y_k$ {\em (Series Dependence)}} &  & \multicolumn{3}{c}{$P^v_k$ {\em (Volatility Dependence)} } &  & \multicolumn{3}{c}{$P^y_k$ {\em (Series Dependence)}} &  & \multicolumn{3}{c}{$P^v_k$ {\em (Volatility Dependence)} } \\
		                                   &   $k=0$ &         &                                   &  &   $k=0$ &         &                                        &  &   $k=0$ &         &                                   &  &   $k=0$ &         &  \\ \cline{2-4}\cline{6-8}\cline{10-12}\cline{14-16}
		\noalign{\smallskip} 
		$JPY_{t}$                          &   1.000 &   0.219 &                             0.338 &  &   1.000 &   0.223 &                                  0.206 &  &   1.000 &   0.211 &                             0.330 &  &   1.000 &   0.216 &                                  0.193 \\
		$AUD_{t}$                          &   0.219 &   1.000 &                             0.574 &  &   0.223 &   1.000 &                                  0.331 &  &   0.211 &   1.000 &                             0.570 &  &   0.216 &   1.000 &                                  0.319 \\
		$EUR_{t}$                          &   0.338 &   0.574 &                             1.000 &  &   0.206 &   0.331 &                                  1.000 &  &   0.330 &   0.570 &                             1.000 &  &   0.193 &   0.319 &                                  1.000 \\
		                                   &   $k=1$ &         &                                   &  &   $k=1$ &         &                                        &  &   $k=1$ &         &                                   &  &   $k=1$ &         &  \\ \cline{2-4}\cline{6-8}\cline{10-12}\cline{14-16}
		\noalign{\smallskip}
		$JPY_{t-1}$                        &  -0.014 &  -0.036 &                            -0.018 &  &   0.058 &   0.077 &                                  0.067 &  &  -0.020 &  -0.030 &                            -0.021 &  &   0.043 &   0.058 &                                  0.044 \\
		$AUD_{t-1}$                        &   0.001 &  -0.017 &                            -0.029 &  &   0.053 &   0.086 &                                  0.073 &  &   0.004 &  -0.013 &                            -0.028 &  &   0.040 &   0.070 &                                  0.057 \\
		$EUR_{t-1}$                        &  -0.013 &  -0.023 &                            -0.019 &  &   0.054 &   0.070 &                                  0.055 &  &  -0.014 &  -0.017 &                            -0.016 &  &   0.040 &   0.059 &                                  0.043 \\ \cline{2-8}\cline{10-16}
		                                                           \multicolumn{8}{c}{BEKK(1,1) model}                                                             &                                         \multicolumn{8}{c}{DCC-GARCH(1,1) model}                                         \\ \cline{2-8}\cline{10-16}
		                                   & $JPY_t$ & $AUD_t$ &                           $EUR_t$ &  & $JPY_t$ & $AUD_t$ &                                $EUR_t$ &  & $JPY_t$ & $AUD_t$ &                           $EUR_t$ &  & $JPY_t$ & $AUD_t$ &                                $EUR_t$ \\ \cline{2-4}\cline{6-8}\cline{10-12}\cline{14-16}
		                                   & \multicolumn{3}{c}{$P^y_k$ {\em (Series Dependence)}} &  & \multicolumn{3}{c}{$P^v_k$ {\em (Volatility Dependence)} } &  & \multicolumn{3}{c}{$P^y_k$ {\em (Series Dependence)}} &  & \multicolumn{3}{c}{$P^v_k$ {\em (Volatility Dependence)} } \\
		                                   &   $k=0$ &         &                                   &  &   $k=0$ &         &                                        &  &   $k=0$ &         &                                   &  &   $k=0$ &         &  \\ \cline{2-4}\cline{6-8}\cline{10-12}\cline{14-16}
		\noalign{\smallskip}
		$JPY_{t}$                          &   1.000 &   0.331 &                             0.419 &  &   1.000 &   0.109 &                                  0.150 &  &   1.000 &   0.088 &                             0.143 &  &   1.000 &   0.036 &                                  0.042 \\
		$AUD_{t}$                          &   0.331 &   1.000 &                             0.449 &  &   0.109 &   1.000 &                                  0.177 &  &   0.088 &   1.000 &                             0.262 &  &   0.036 &   1.000 &                                  0.077 \\
		$EUR_{t}$                          &   0.419 &   0.449 &                             1.000 &  &   0.150 &   0.177 &                                  1.000 &  &   0.143 &   0.262 &                             1.000 &  &   0.042 &   0.077 &                                  1.000 \\
		                                   &   $k=1$ &         &                                   &  &   $k=1$ &         &                                        &  &   $k=1$ &         &                                   &  &   $k=1$ &         &  \\ \cline{2-4}\cline{6-8}\cline{10-12}\cline{14-16}
		\noalign{\smallskip}
		$JPY_{t-1}$                        &       0 &       0 &                                 0 &  &   0.047 &   0.009 &                                  0.009 &  &       0 &       0 &                                 0 &  &   0.047 &   0.002 &                                  0.001 \\
		$AUD_{t-1}$                        &       0 &       0 &                                 0 &  &   0.006 &   0.088 &                                  0.016 &  &       0 &       0 &                                 0 &  &   0.001 &   0.115 &                                  0.009 \\
		$EUR_{t-1}$                        &       0 &       0 &                                 0 &  &   0.010 &   0.017 &                                  0.061 &  &       0 &       0 &                                 0 &  &       0 &   0.010 &                                  0.087 \\ \bottomrule
	\end{tabular}
	\caption{Matrices of pairwise Spearman correlations in the series ($P^y_k$; left-hand side) and the 
		volatility ($P^v_k$; right-hand side) for the three-dimensional exchange rate example. Results are reported separately
		for the two fitted copula models, and the 
		BEKK and DCC models. 
		The daily returns on the
		USD/JPY, USD/AUD and USD/EUR exchange rates are denoted at time $t$ as $JPY_t, AUD_t$ and $EUR_t$, respectively. 
		For each model, the top 
		rows give cross-sectional
		dependence ($k=0$), and the bottom rows give first order serial dependence ($k=1$). For example, for
		the time series copula model with Copula A,
		the pairwise Spearman correlation 
		between $JPY_{t-1}$ and $EUR_t$ is $\rho^y_{JPY,EUR,1}=-0.018$, and the
		corresponding volatility spillover is $\rho^v_{JPY,EUR,1}=0.067$.} 
	\label{tab:multirhos}
\end{table}
\end{landscape}

\begin{table}[H]
	\centering
	\begin{tabular}{lcccccc}
		\toprule
		                          &           \multicolumn{6}{c}{Quantile $\alpha$}            \\
		\multirow{1}{*}{Model}    &   1\%   &  5\%   &   10\%   &   90\%   &  95\%   &  99\%   \\ \midrule
		Copula~A                & 0.82\%  & 4.47\% &  9.54\%  & 89.56\%  & 95.12\% & 99.32\% \\
		Copula~B               & 0.65\%  & 4.36\% &  9.51\%  & 89.50\%  & 94.85\% & 99.24\% \\
		BEKK(1,0)                 & 1.50\%* & 4.63\% & 8.15\%** & 90.95\%* & 95.26\% & 98.77\% \\
		BEKK(1,1)                 & 1.28\%  & 4.63\% & 8.37\%** & 90.51\%* & 94.98\% & 98.96\% \\
		DCC-GARCH(1,1)            & 1.34\%  & 4.77\% & 8.51\%** & 90.27\%* & 94.93\% & 98.88\% \\ \bottomrule
	\end{tabular}
	\caption{Mean exceedances $\hat \alpha$ (in percent)
		over $T-1=3668$ days of 
		one day ahead VaR forecasts of returns on an equally-weighted currency portfolio. Results
		are given for five multivariate models and six different quantile values.
		Rejection of the null hypothesis of the conditional coverage Christoffersen~(1998) test is denoted with `*' and `**' at the 95\% and 99\% level of 
		confidence, respectively.}
	\label{tab:exceedmulti}
\end{table}

%% file: figs.tex
 \begin{figure}[H]
    	\begin{center}
    		\includegraphics[scale =0.9]{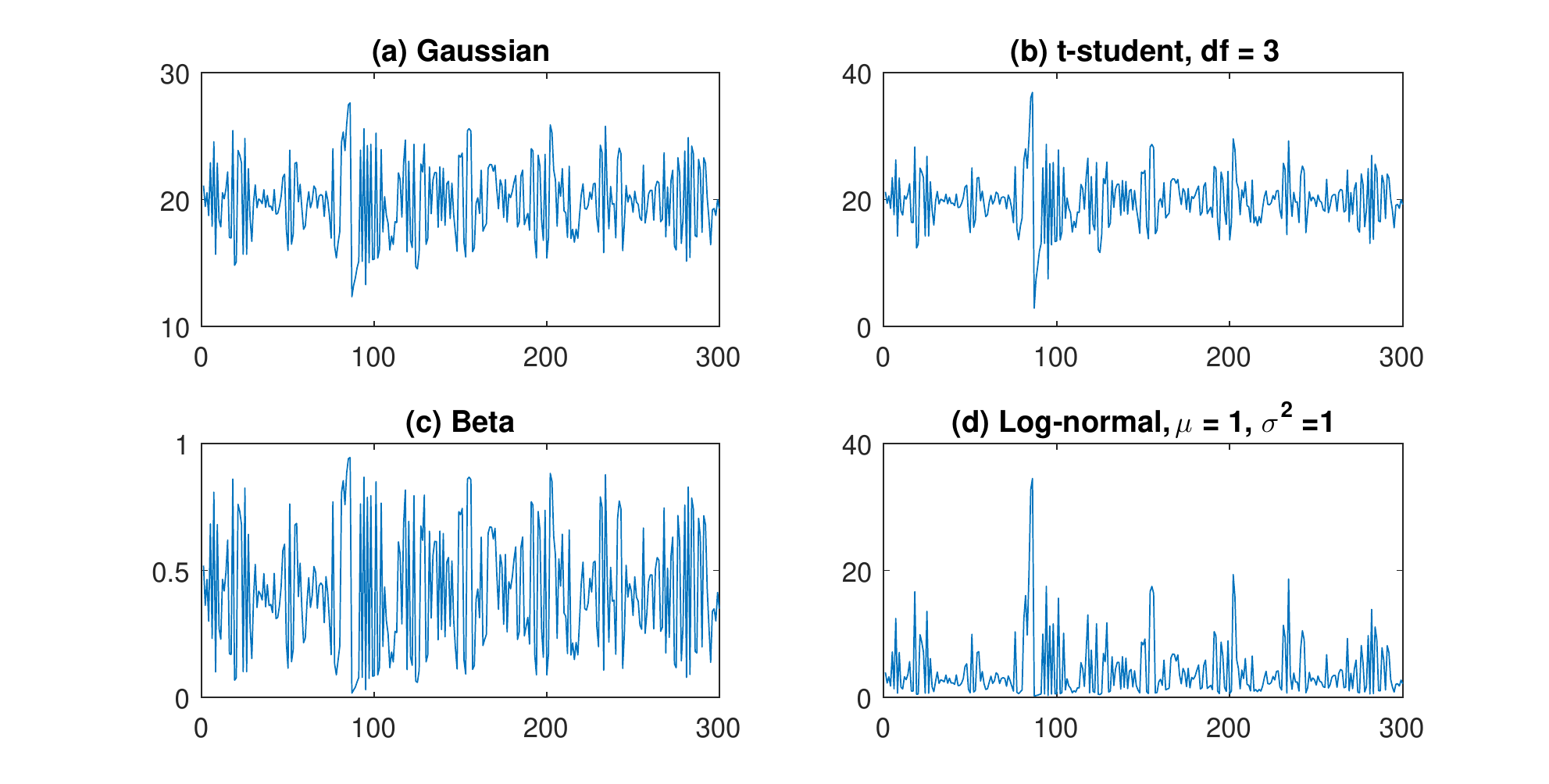}
    	\end{center}
    	\caption{Four heteroskedastic series simulated from the time series copula model.
A vector $\bm{u}$ of length $T=300$ is simulated from 
the copula at Equation~(\ref{eq:coplike}) with
$c_2=c^{MIX}$,
where $w=0.5$, and $c_a,c_b$ are identical t copula densities with parameters
$(\zeta=0.9,\nu=3)$. We then compute $y_t=F^{-1}(u_t)$ 
using the quantile function $F^{-1}$ of the four time invariant marginal 
distributions: (a)~$Y_t\sim N(20,3)$,
(b)~$Y_t\sim 20+3t_3$, (c)~$Y_t\sim Beta(1.5,2)$ and (d)~$\ln(Y_t)\sim N(1,1)$.
}
    	\label{Fig:Persistent_processes}
    \end{figure}

\begin{figure}[H]
	\begin{center}
		\includegraphics[scale =1.2]{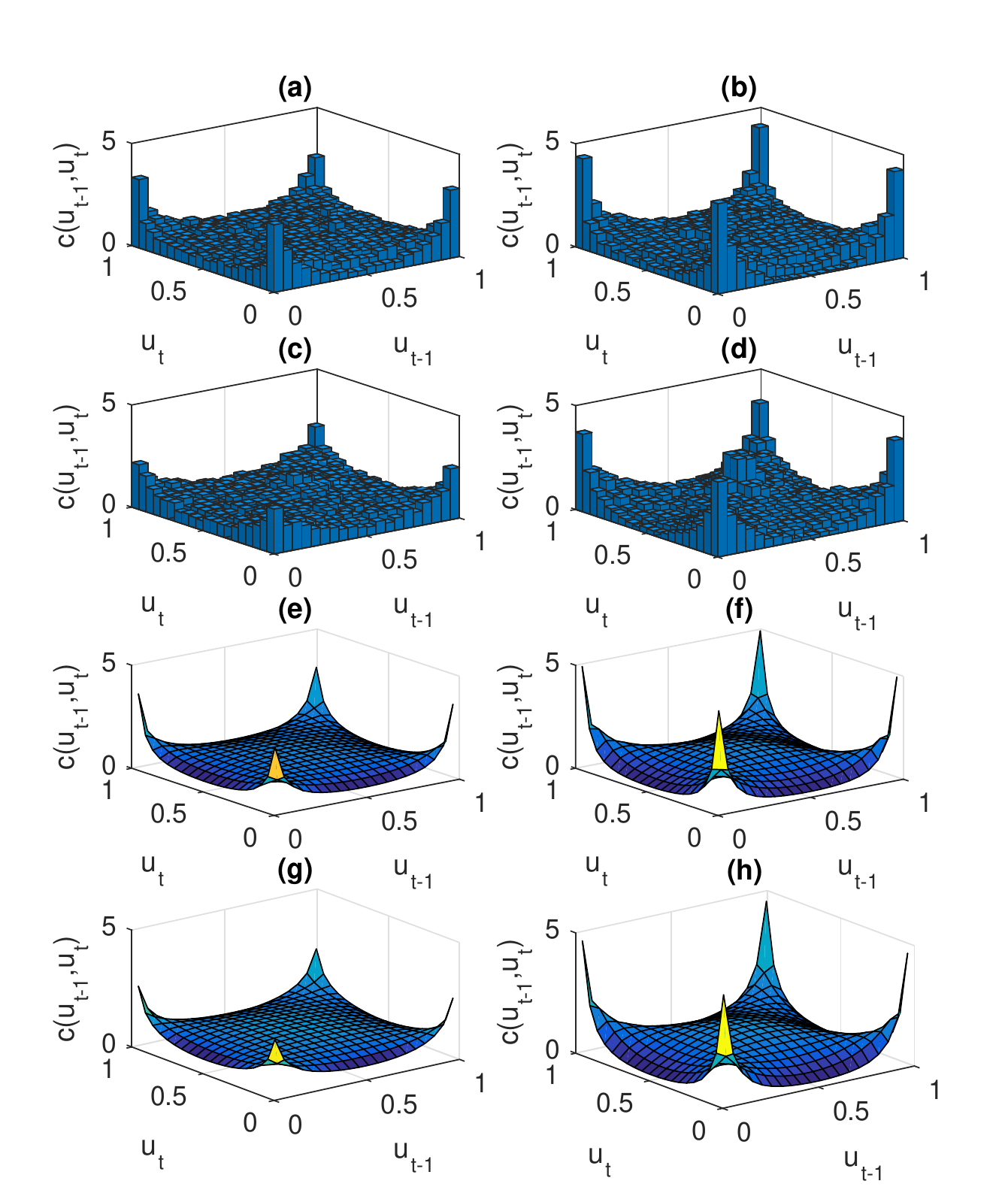}
	\end{center}
	\caption{Copula densities $c_2$ for heteroskedastic time series. 
Panels (a) and (b) present the empirical copula densities
for the ARCH(1) processes with $\alpha_1 = 0.5$ and $\alpha_1 = 0.9$, respectively. Panels (c) and (d) display the empirical density copulas for the SV(1) processes with $\phi_1 = 0.5$ and $\phi_1 = 0.9$, respectively.
Panels (e) to (h) plot the mixture copulas $c^{MIX}$ (with t-copula components) fitted to the same data used to compute the
densities in panels (a) to (d), respectively.} 
	\label{Fig:Mixture_ofCop_densities}
\end{figure}

\begin{figure}[H]
	\begin{center}
		\includegraphics[scale =0.8]{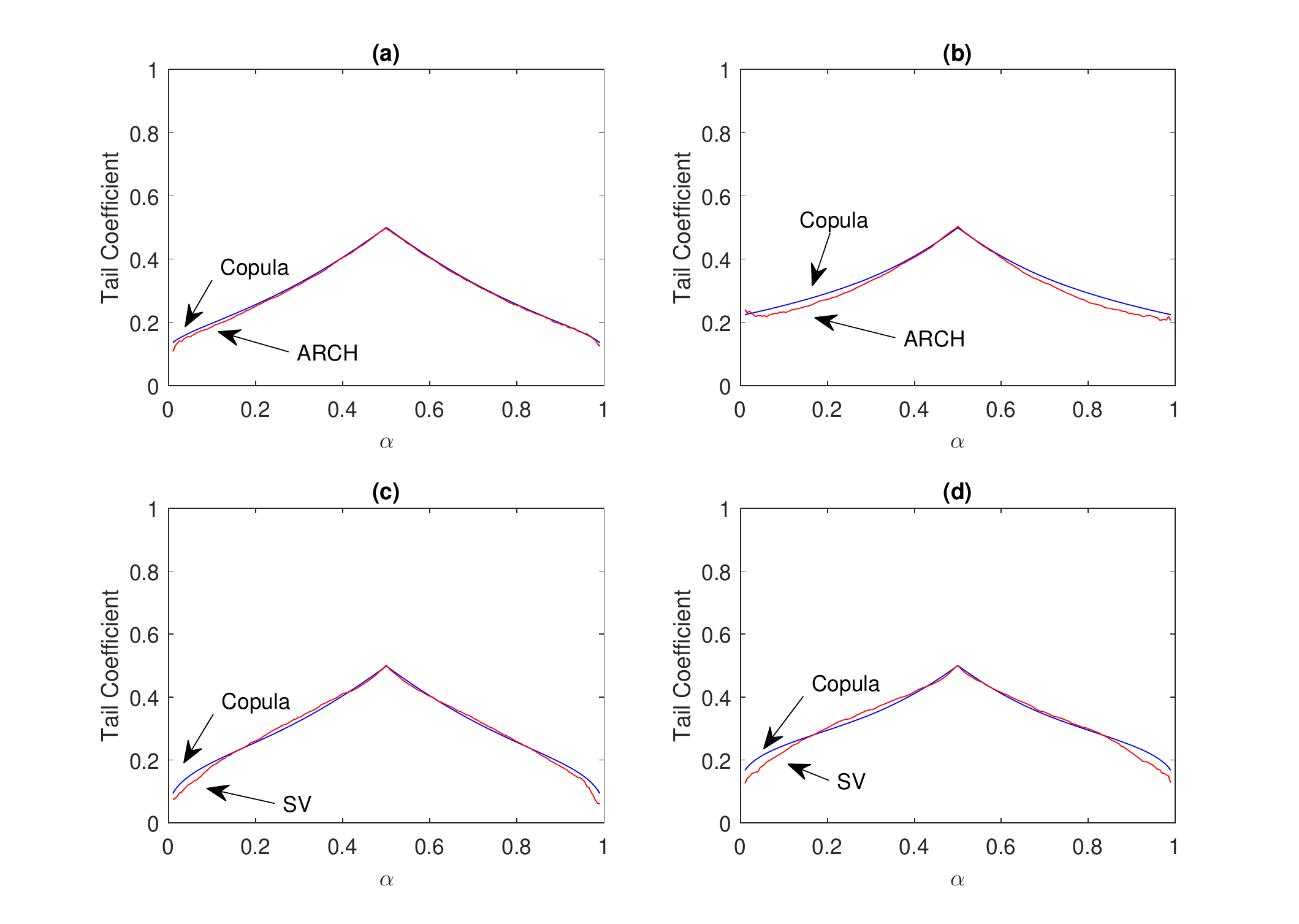}
	\end{center}
	\caption{Quantile dependence coefficients for ARCH(1) and SV(1) models. Panels (a) and (b) display the 
		quantile coefficients for the ARCH(1) processes with $\alpha_1 = 0.5$ and $\alpha_1 = 0.9$, respectively. Panels (c) and (d) show the quantile coefficients for the SV(1) processes with $\phi_1 = 0.5$ and $\phi_1 = 0.9$.
		The coefficient $\lambda_{\mbox{\tiny low}}(\alpha)$ is plotted for $\alpha<0.5$, and 
		$\lambda_{\mbox{\tiny up}}(\alpha)$ for $\alpha>0.5$.
The red line gives the empirical quantile coefficients, and the blue line that from the copula model.}
	\label{Fig:Tal_dep_arch1}
\end{figure}

  \begin{figure}[H]
  	\begin{center}
  		\includegraphics[scale =0.7]{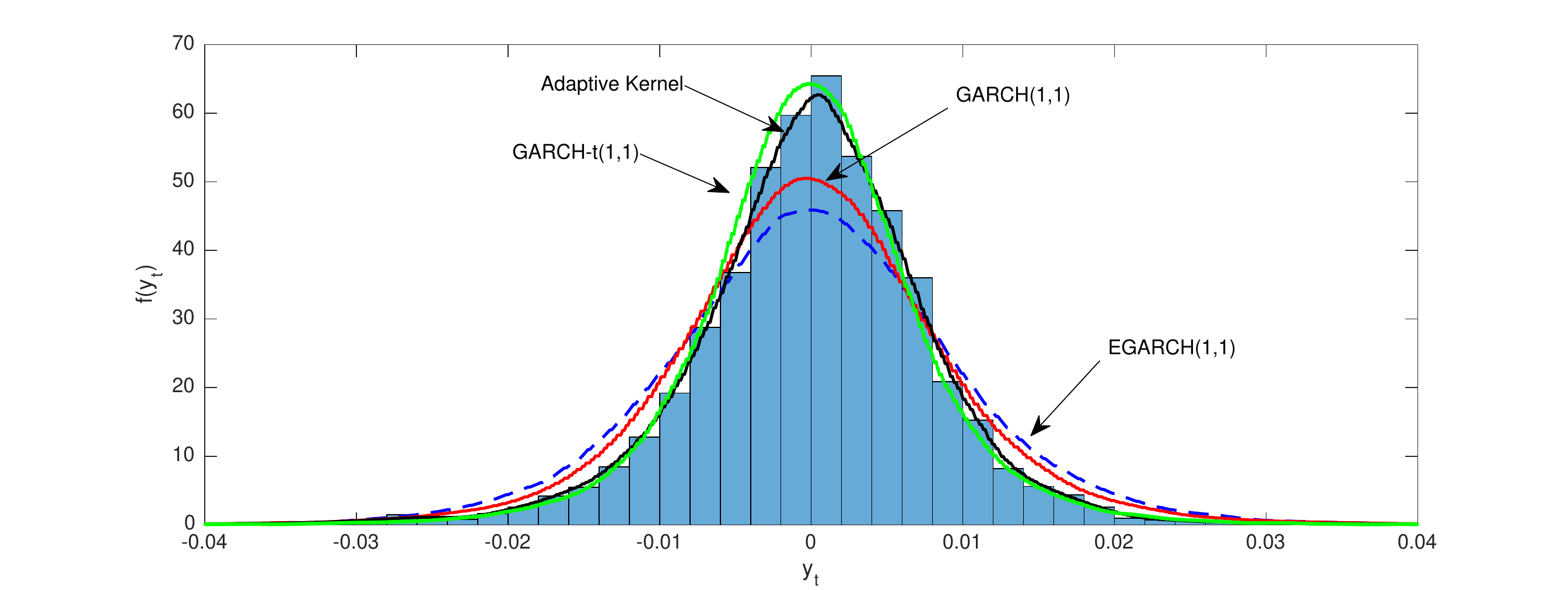}
  	\end{center}
  	\caption{Histogram of daily USD/AUD returns, the locally adaptive kernel density estimate (black line),
		and the marginal distributions of fitted EGARCH(1,1) (blue dashed line), GARCH(1,1) (red line)
	and GARCH-t(1,1) (green line) models.}
  	 	\label{Fig:AUD_histogram}
  \end{figure}

  \begin{figure}[H]
  	\begin{center}
  		\includegraphics[scale =1]{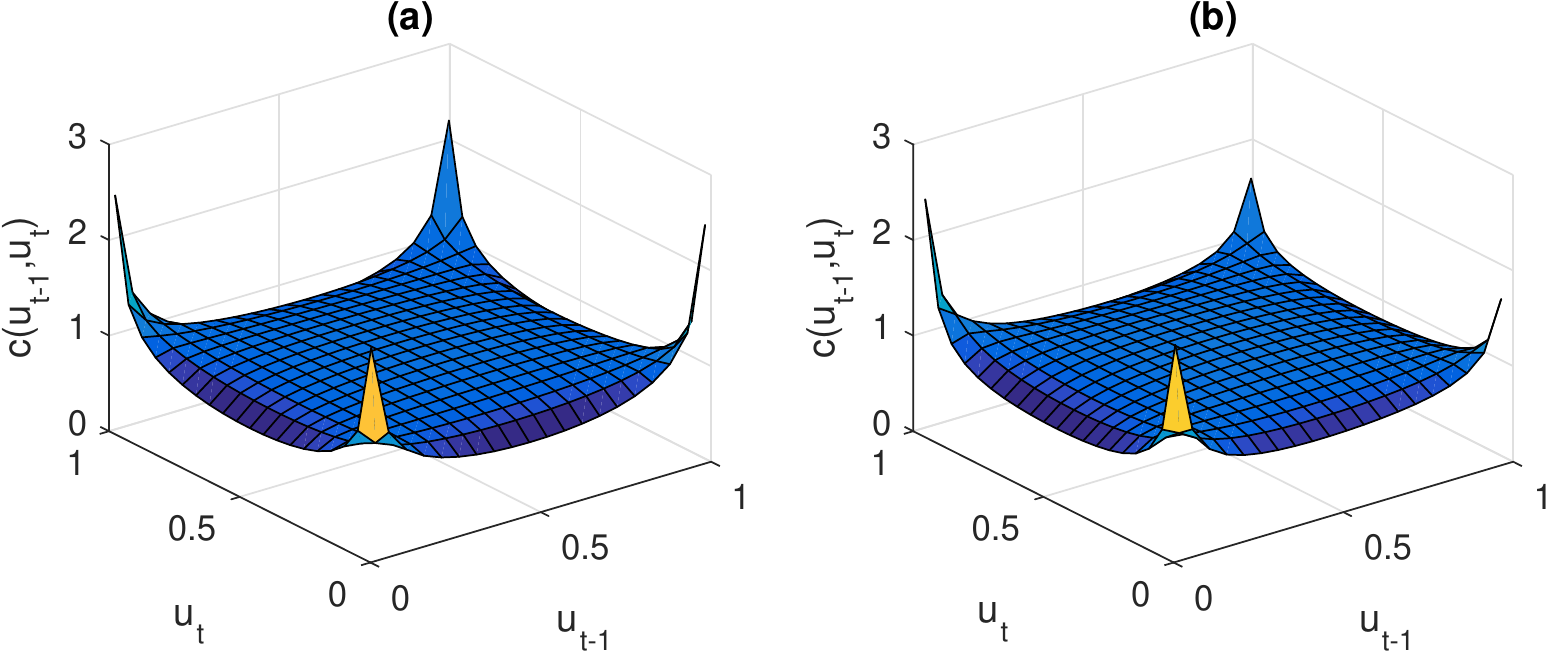}
  	\end{center}
	\caption{Bivariate copula densities $c_2$ fitted to the USD/AUD returns. A mixture copula was 
	used with t-copula components in panel~(a), and convex Gumbel components in panel~(b).}
  	 	\label{fig:AUD_USD_fittedcop}
  \end{figure}
 
\begin{figure}[H]
	\begin{center}
		\includegraphics[scale = 0.6]{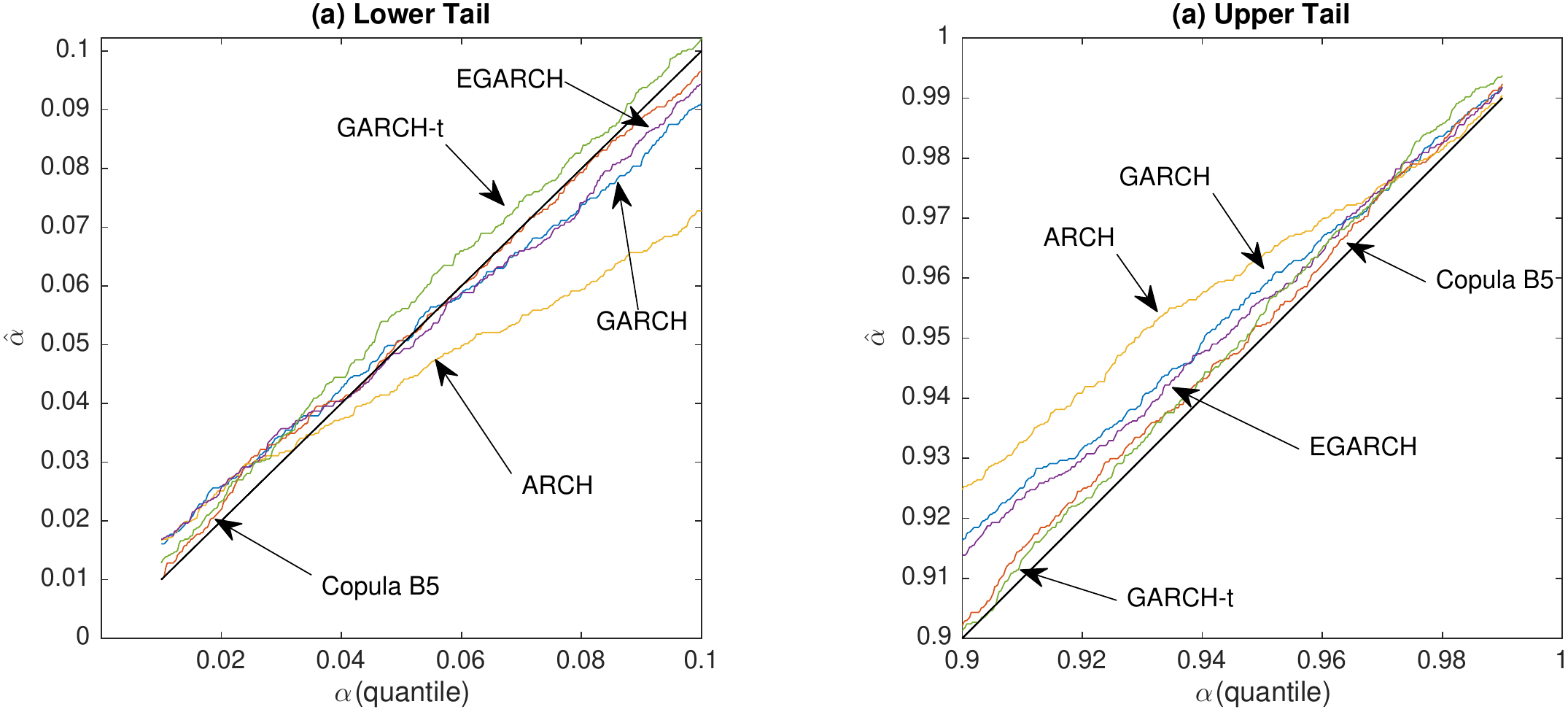}
	\end{center}
	\caption{Coverage results for one day ahead VaR predictions for 
USD/AUD daily returns. 
Panel~(a) displays results for the lower tail, and panel~(b) for the upper tail.
Results are given for the ARCH(1), GARCH(1,1), EGARCH(1,1), GARCH-t(1,1) and
Copula~B5 models.}
	\label{Fig:Back_testing_AUD}
\end{figure}
  
\begin{figure}[H]
	\begin{center}
		\includegraphics[scale = 0.6]{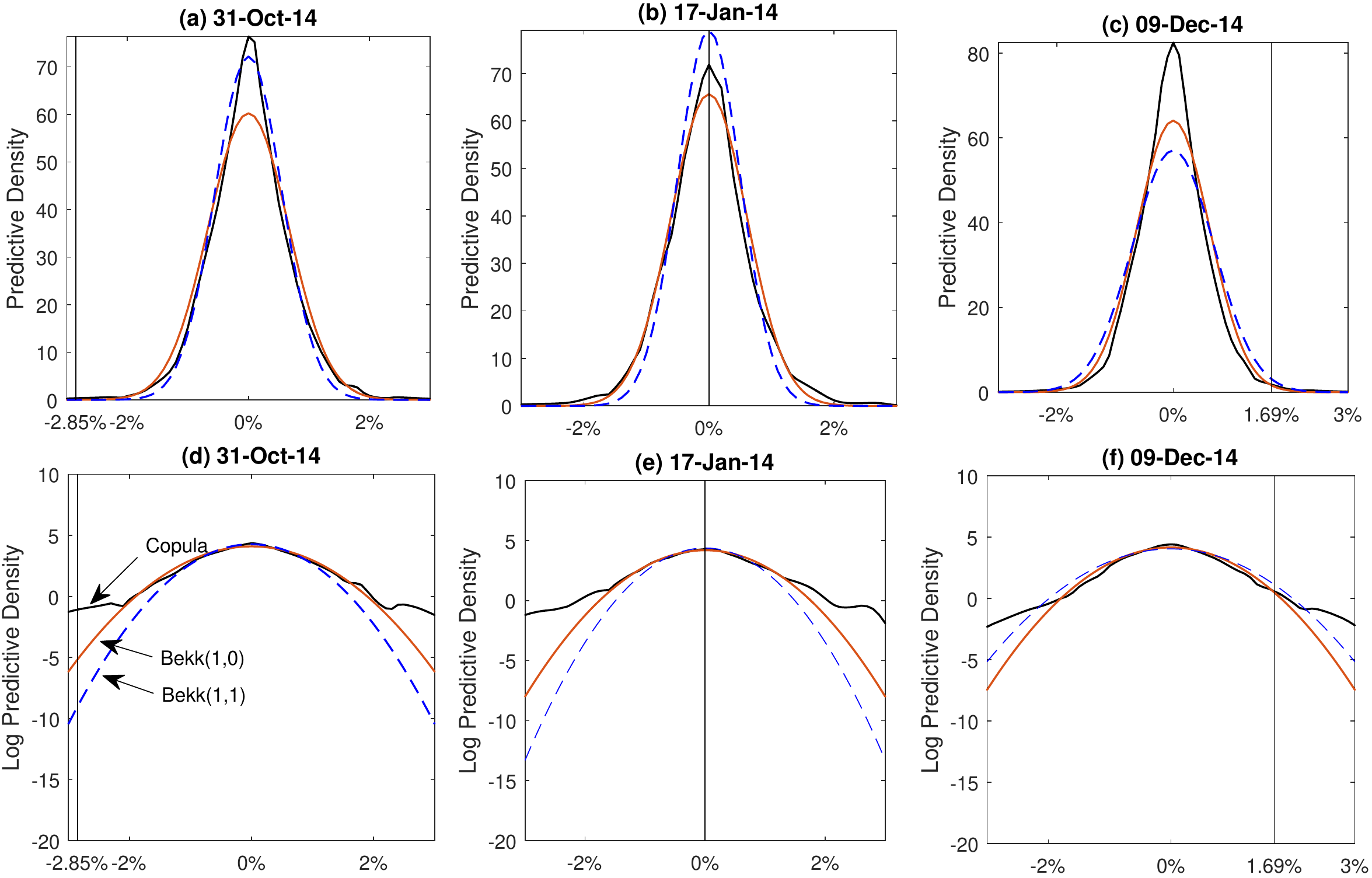}
	\end{center}
	\caption{Panels~(a--c) provide one day ahead forecast densities for the USD/JPY return
		from the Copula~A (black line), BEKK(1,0) (red line) and BEKK(1,1) models (blue dashed line), 
respectively. Panels~(d--e) plot the logarithm of the same three forecast densities, to aid
comparison of the tails.
Forecasts are made for three specific days: (a,d) 31 Oct. 2014; (b,e) 17 Jan. 2014; and, (c,f) 9 Dec. 2014.
These days have the smallest, largest and `most close to zero' returns observed during the last 
18 months of the series. 
Vertical lines plot these observed values.}
		\label{fig:fxdenfore}
\end{figure}

%% file: web-append.tex
\noindent
{\bf \Large{Online Appendix for `Time Series Copulas for Heteroskedastic Data'
}}
\vspace{10pt}

\noindent
Part~A of this online appendix provides a comparison of the vine copula and an ARCH(3) model,
while Part~B provides some additional output from the empirical analysis in the manuscript.
\newpage
\setcounter{figure}{0}
\setcounter{table}{0}

\begin{figure}[H]
	\begin{center}
		\includegraphics[scale =0.85]{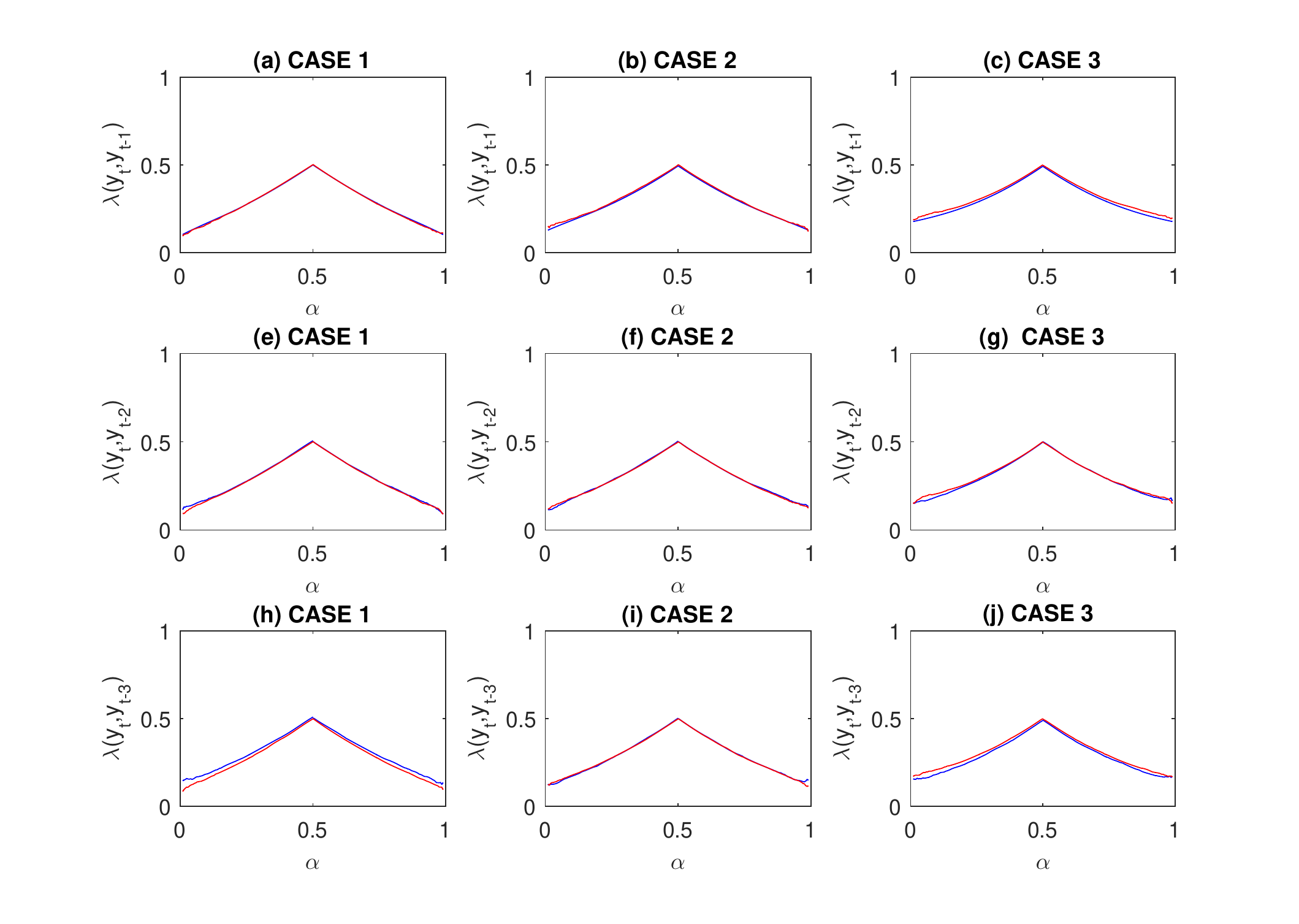}
	\end{center}
	\caption{Quantile dependence coefficients for the ARCH(3) model. Rows 1 to 3 show the quantile coefficients for the bivariate marginal copulas 
$c(u_t,u_{t-1})$, $c(u_t,u_{t-2})$ and $c(u_t,u_{t-3})$, respectively, while columns 1 to 3 indicate the ARCH(3) parameters case. For each panel, the red line shows the empirical quantile coefficient, and the blue line shows the copula 
fitted quantile coefficient.}
	\label{Fig:Tal_dep_arch3}
\end{figure}

\noindent {\bf \large{Part~A: Comparision of Vine Copula and ARCH(3) Model}}\\
\noindent
Here, we illustrate that the vine copula in Equation~(\ref{eq:DvineLikeli})
can replicate the dependence characteristics of three ARCH(3) models.
The ARCH(3) model
has conditional variance
$\sigma^2_t=\alpha_0+\sum_{j=1}^3 \alpha_j y_{t-j}^2$. We consider
three cases, where $(\alpha_0,\alpha_1,\alpha_2,\alpha_3)$ equals
(i)~(0.01,0.2,0.2,0.2), (ii)~(0.01,0.3,0.2,0.2), and (iii)~(0.01,0.5,0.2,0.2).
We simulate $T=50,000$ iterates from each case, and fit the D-vine copula
by maximizing $c^{DV}$, which is the 
likelihood conditional on the copula data. The adequacy of the fitted
vine is measured by considering the bivariate marginal copulas with densities
$c(u_{t-k},u_{t})$, for $k\geq 1$,
which capture pairwise serial dependence.
Figure~\ref{Fig:Tal_dep_arch3} plots the quantile dependence
coefficients of the marginal copulas from the D-vine with $k=1,2$ \& 3, when fit
to each of the three datasets. Also plotted are the corresponding empirical 
quantile coefficients. 
In 
all three cases, the
quantile dependence between $y_t$ and its three lagged values for the D-vine models are
close to the corresponding empirical values. While not presented here,
this is also true for the quantile dependence coefficients of the other two 
quadrants.

\newpage 

\noindent {\bf \large{Part~B: Additional Empirical Output}\\
\noindent

\begin{table}[H]
	\centering
	\begin{tabular}{llrrrrr}
		\toprule
		     & Model                                       & $\zeta^a>0$ & $\nu^a$ & $\zeta^b>0$ & $\nu^b$ &   $w$ \\ \midrule
		ARCH & $(\alpha_0 = 0.01, \alpha_1 = 0.5)$         &       0.705 &  39.996 &       0.179 &   2.984 & 0.191 \\
		     & $(\alpha_0 = 0.01, \alpha_1 = 0.9)$         &       0.678 &   6.004 &       0.701 &   7.044 & 0.509 \\
		SV   & $(\phi_1 = 0.5, \sigma^2=2.5, \bar{h}=0.8)$ &       0.454 &  20.019 &       0.555 &  39.994 & 0.547 \\
		     & $(\phi_1 = 0.9, \sigma^2=2, \bar{h}=0.8)$   &       0.693 &  10.740 &       0.728 &  15.031 & 0.512 \\ \bottomrule
	\end{tabular}
	\caption{MLE parameter estimates for mixture of t copulas fitted to ARCH(1) and SV(1) models.}
	\label{Tab:copulaparARCH1SV1}
\end{table}

\begin{figure}[h]
	\begin{center}
		\includegraphics[scale =0.8]{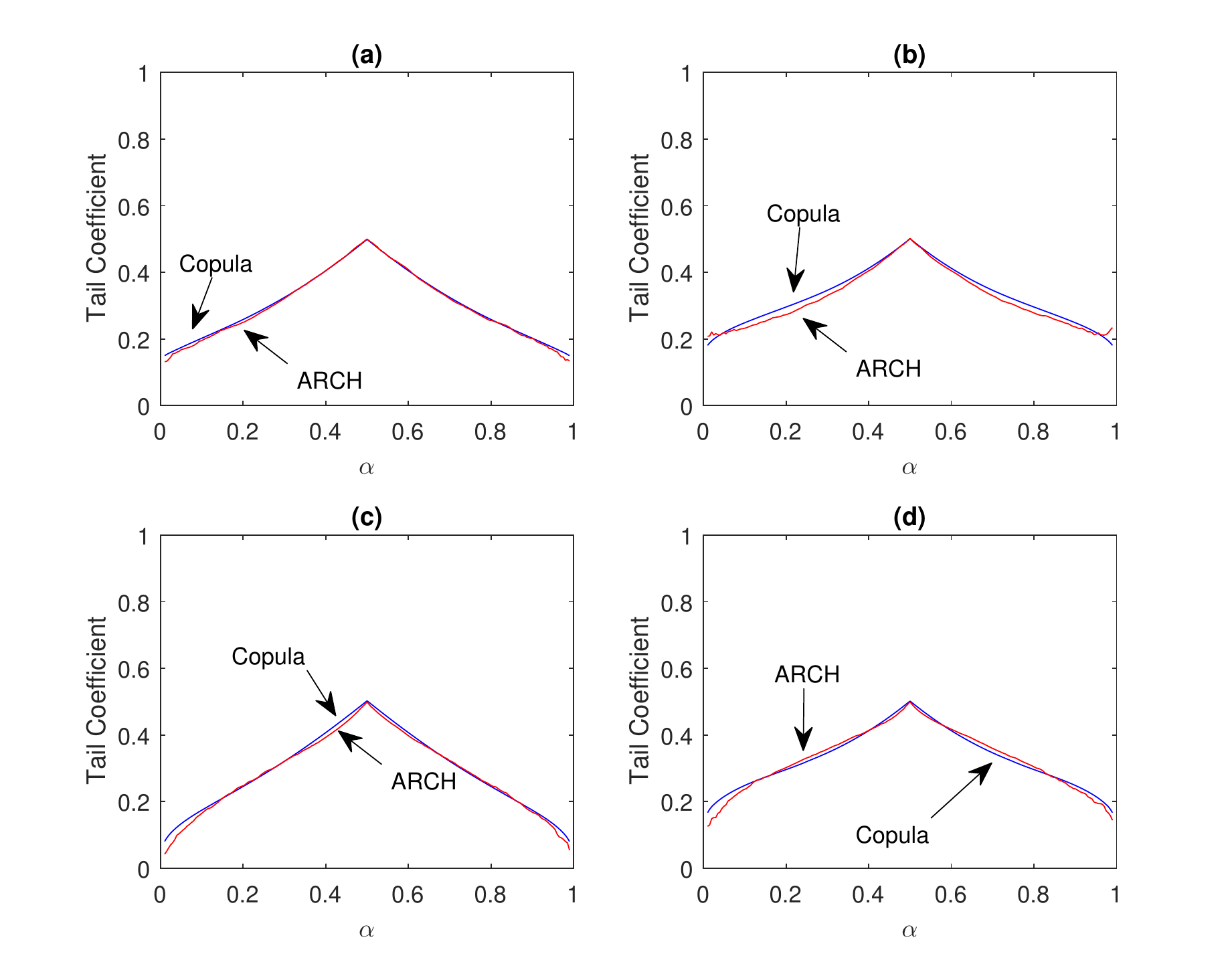}
	\end{center}
	\caption{The off-diagonal quantile dependence coefficients
		$\lambda_{LU}(\alpha)\equiv \mbox{Pr}(u_t>(1-\alpha)|u_{t-1}<\alpha)$ and
		$\lambda_{UL}(\alpha)\equiv \mbox{Pr}(u_t<(1-\alpha)|u_{t-1}>\alpha)$
		for the ARCH(1) and SV(1) models in Section~\ref{sec:2.1}. Panels (a) and (b) display the 
		quantile coefficients for the ARCH(1) processes with $\alpha_1 = 0.5$ and $\alpha_1 = 0.9$, respectively. Panels (c) and (d) show the quantile coefficients for the SV(1) processes with $\phi_1 = 0.5$ and $\phi_1 = 0.9$. The coefficient $\lambda_{LU}(\alpha)$ is plotted to left of $\alpha = 0.5$, and
		$\lambda_{UL}(\alpha)$ to the right.
The red line gives the empirical quantile coefficients, and the blue line that from the copula model.}
	\label{oafig:offdiag}
\end{figure}

\begin{figure}[H]
	\begin{center}
		\includegraphics[scale =1]{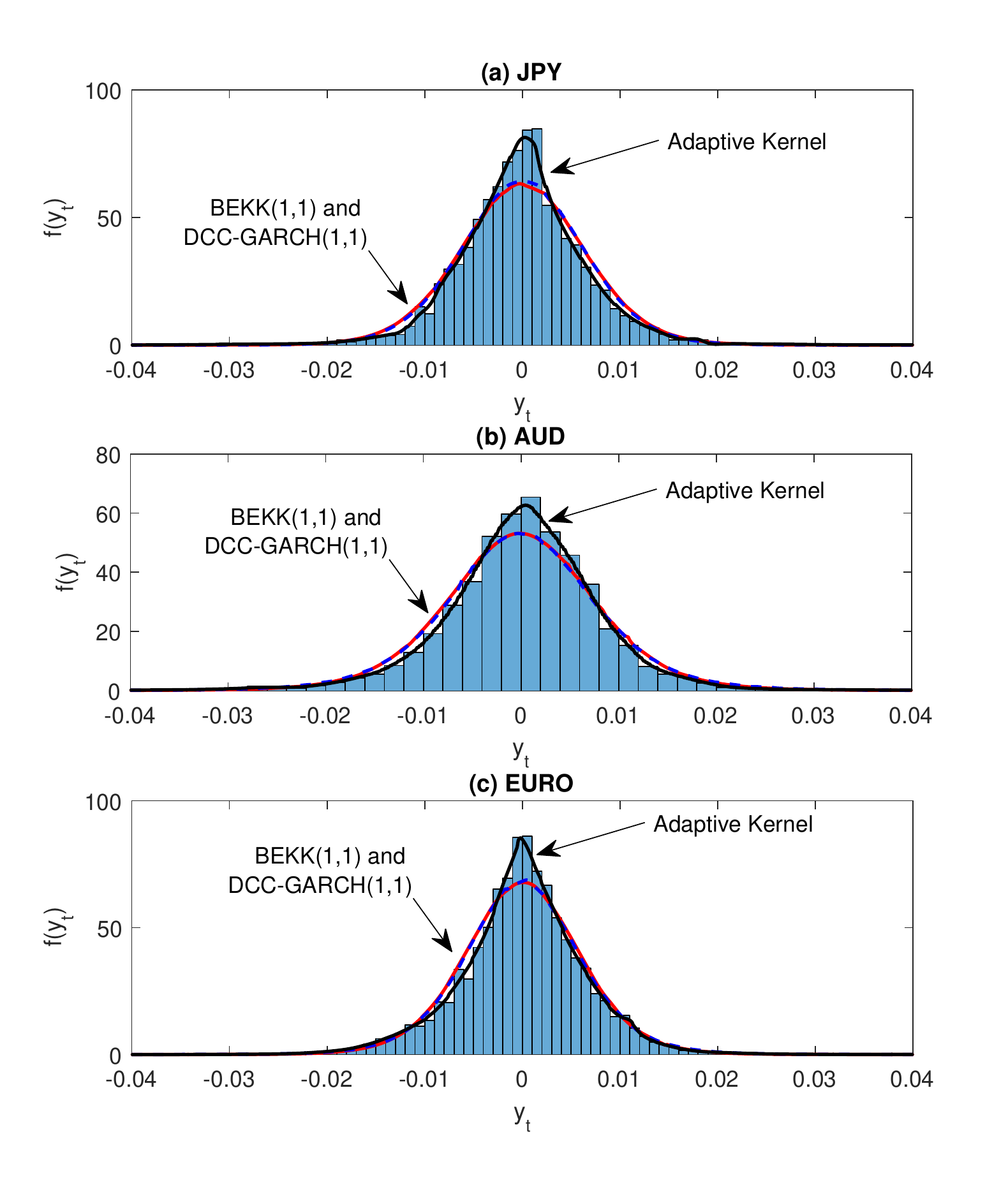}
	\end{center}
   \caption{Histograms of daily foreign exchange returns on (a)~USD/JPY, 
(b)~USD/AUD and (c)~USD/EUR. Also plotted are the locally adaptive 
kernel density estimates (black lines), and the marginal distributions
of the fitted BEKK(1,1) (red lines) and DCC-GARCH(1,1) (blue dashed lines) models.}
	\label{fig:3hist}
\end{figure}

\begin{figure}[H]
	\begin{center}
		\includegraphics[scale =1]{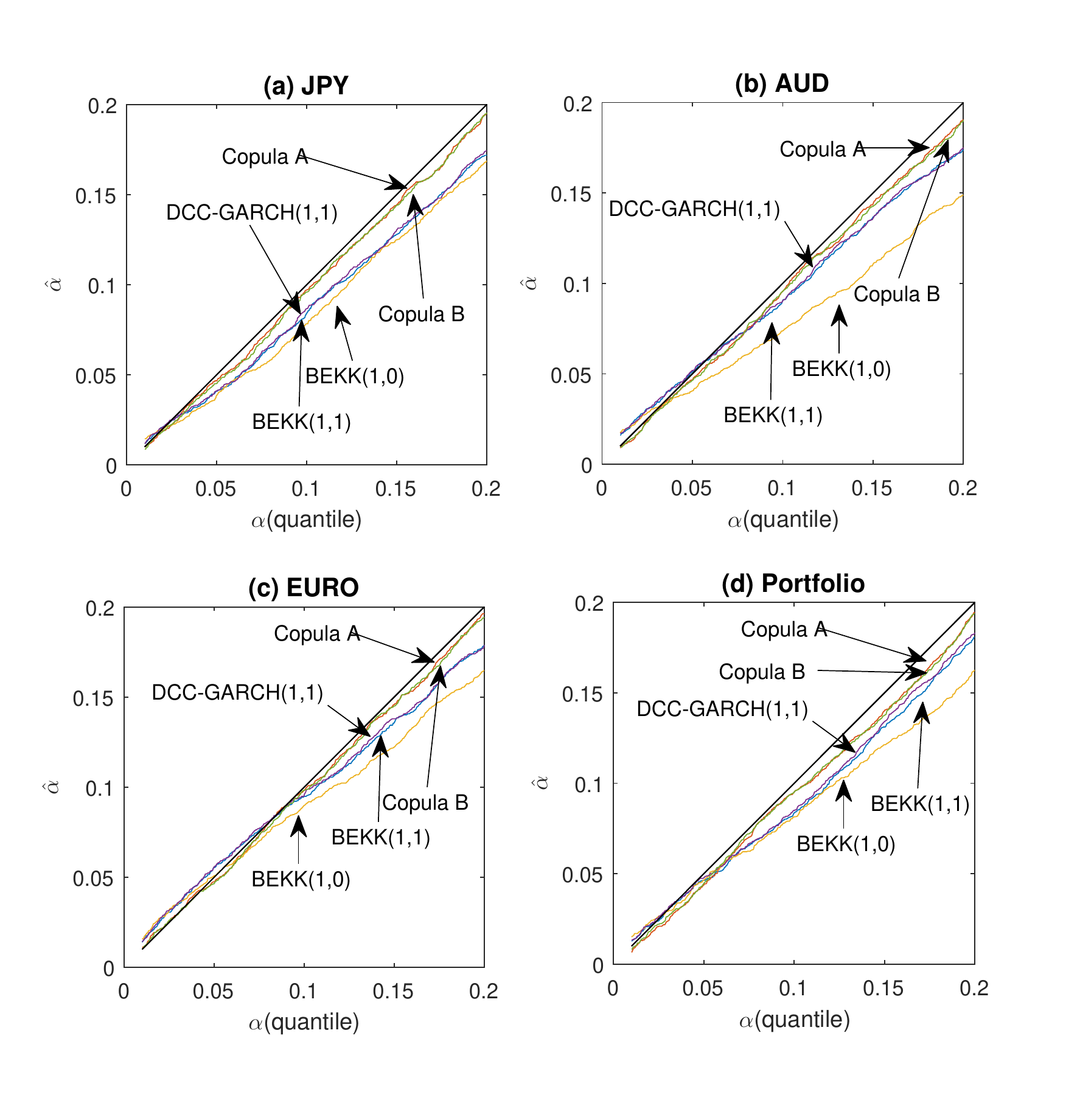}
	\end{center}
	\caption{One-step ahead coverage results for lower quantile predictions for the multivariate exchange rate application.
		Each line corresponds to a different model, with lines closer to 45 degree
	indicating more accurate VaR forecasts.}
		\label{Fig:Back_testing_currencies_low}
\end{figure}

\begin{figure}[H]
	\begin{center}
		\includegraphics[scale = 1]{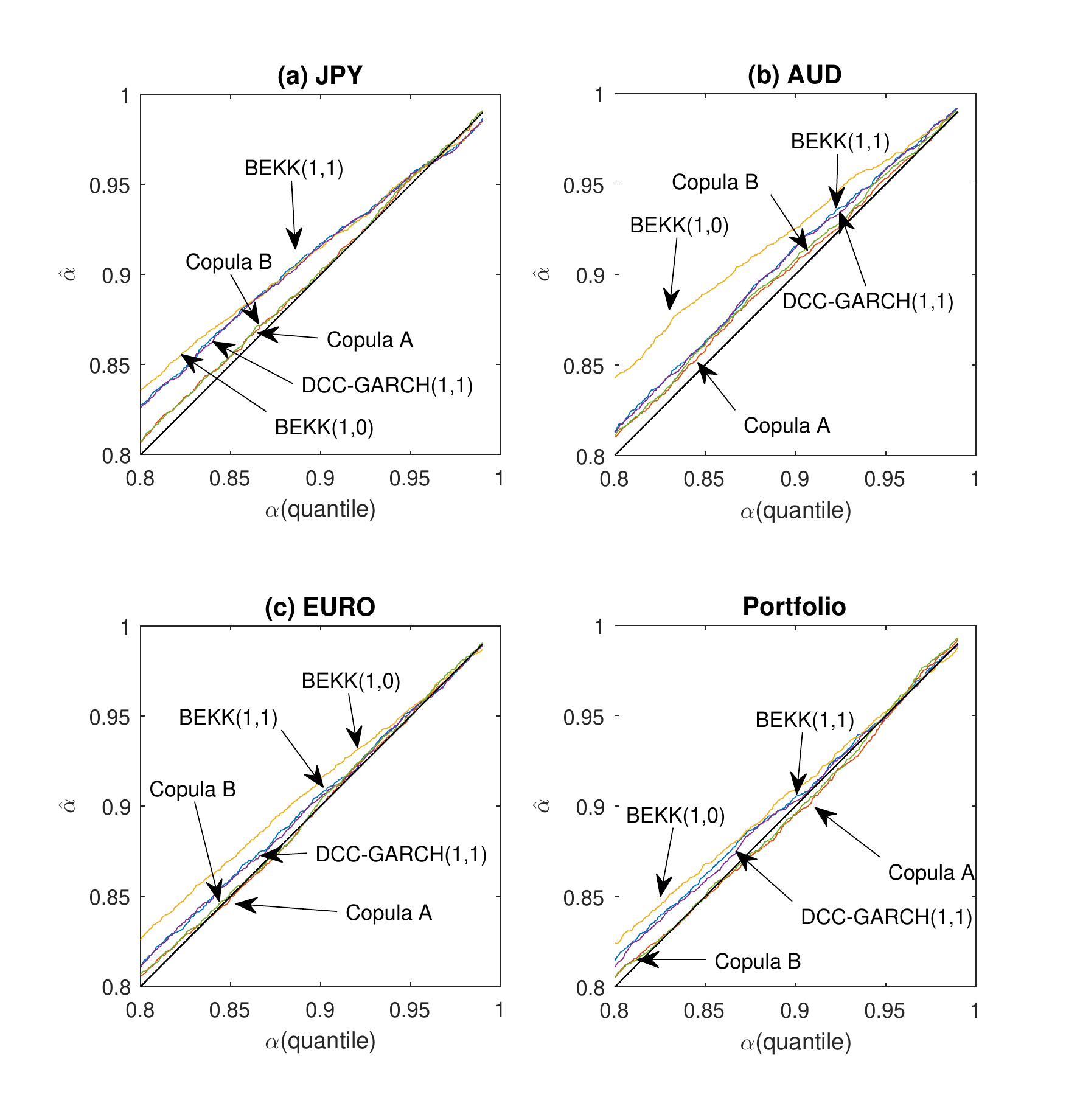}
	\end{center}
	\caption{One-step ahead coverage results for upper quantile predictions for the multivariate exchange rate application.
		Each line corresponds to a different model, with lines closer to 45 degree
	indicating more accurate VaR forecasts.}
		\label{Fig:Back_testing_currencies_upp}
\end{figure}

%% file: main.bbl
\begin{thebibliography}{}

\bibitem[Aas et al. (2009)]{Aas2009182}
Aas, K., Czado, C., Frigessi, A., and Bakken, H. (2009).
\newblock {P}air-copula constructions of multiple dependence.
\newblock {\em {I}nsurance: {M}athematics and {E}conomics}, 44(2):182 -- 198.

\bibitem[Almeida and Czado(2012)]{almeida2012}
Almeida, C. and Czado, C. (2012).
\newblock {E}fficient {{B}ayesian} inference for stochastic time-varying copula
  models.
\newblock {\em {C}omputational {S}tatistics \& {D}ata {A}nalysis},
  56(6):1511--1527.

\bibitem[Baillie and Bollerslev(1991)]{baillie1991}
Baillie, R.~T. and Bollerslev, T. (1991).
\newblock {I}ntra-day and inter-market volatility in foreign exchange rates.
\newblock {\em {T}he {R}eview of {E}conomic {S}tudies}, 58(3):565--585.

\bibitem[Baillie and Bollerslev(2002)]{baillie2002}
Baillie, R.~T. and Bollerslev, T. (2002).
\newblock {T}he message in daily exchange rates: a conditional-variance tale.
\newblock {\em {J}ournal of {B}usiness \& {E}conomic {S}tatistics},
  20(1):60--68.

\bibitem[Beare(2010)]{beare2010}
Beare, B.~K. (2010).
\newblock {C}opulas and {T}emporal {D}ependence.
\newblock {\em {E}conometrica}, 78(1):395--410.

\bibitem[Beare(2012)]{beare2012}
Beare, B.~K. (2012).
\newblock {A}rchimedean copulas and temporal dependence.
\newblock {\em {E}conometric {T}heory}, 28(06):1165--1185.

\bibitem[Beare and Seo(2015)]{beare15}
Beare, B.~K. and Seo, J. (2015).
\newblock {V}ine copula specifications for stationary multivariate {M}arkov
  chains.
\newblock {\em {J}ournal of {T}ime {S}eries {A}nalysis}, 36(2):228--246.

\bibitem[Biller(2009)]{biller2009}
Biller, B. (2009).
\newblock Copula-based multivariate input models for stochastic simulation.
\newblock {\em {O}perations {R}esearch}, 57(4):878--892.

\bibitem[Biller and Nelson(2003)]{biller2003}
Biller, B. and Nelson, B. (2003).
\newblock {M}odeling and generating multivariate time-series input processes
  using a vector autoregressive technique.
\newblock {\em {A}{C}{M} {T}ransactions on {M}odeling and {C}omputer
  {S}imulation}, 13(3):1049--3301.

\bibitem[Boothe and Glassman(1987)]{boothe1987}
{B}oothe, P. and {G}lassman, D. (1987).
\newblock {T}he statistical distribution of exchange rates: empirical evidence
  and economic implications.
\newblock {\em {J}ournal of {I}nternational {E}conomics}, 22(3):297--319.

\bibitem[Brechmann and Czado(2015)]{brechmann15}
Brechmann, E.~C. and Czado, C. (2015).
\newblock {COPAR}-- multivariate time series modeling using the copula
  autoregressive model.
\newblock {\em {A}pplied {S}tochastic {M}odels in {B}usiness and {I}ndustry},
  31(4):495--514.

\bibitem[Brockwell and Davis(1991)]{Brockwell1991}
Brockwell, P.~J. and Davis, R.~A. (1991).
\newblock {\em {T}ime {S}eries : {T}heory and {M}ethods}.
\newblock {S}pringer {S}eries in {S}tatistics. {S}pringer, {N}ew {Y}ork
  ({N}.{Y}.).

\bibitem[Celeux et~al.(2006)]{celeux2006}
Celeux, G., Forbes, F., Robert, C.~P., and Titterington, D.~M. (2006).
\newblock Deviance information criteria for missing data models.
\newblock {\em Bayesian analysis}, 1(4):651--673.

\bibitem[Chen and Fan(2006)]{chen2006}
Chen, X. and Fan, Y. (2006).
\newblock Estimation of copula-based semiparametric time series models.
\newblock {\em Journal of Econometrics}, 130(2):307--335.

\bibitem[Chen et al.(2009)]{chen2009}
Chen, X., Wu, W.~B., and Yi, Y. (2009).
\newblock {E}fficient {E}stimation of {C}opula-based {S}emiparametric {M}arkov
  {M}odels.
\newblock {\em Annals of statistics}, 37(6B):4214--4253.

\bibitem[Christoffersen(1998)]{christoff1998}
Christoffersen, P.~F. (1998).
\newblock Evaluating interval forecasts.
\newblock {\em International Economic Review}, 39:841--862.

\bibitem[Clark and Ravazzolo(2015)]{clark2015}
Clark, T.~E. and Ravazzolo, F. (2015).
\newblock {M}acroeconomic forecasting performance under alternative
  specifications of time-varying volatility.
\newblock {\em Journal of Applied Econometrics}, 30(4):551--575.

\bibitem[Creal and Tsay(2015)]{creal2015}
Creal, D.~D. and Tsay, R.~S. (2015).
\newblock High dimensional dynamic stochastic copula models.
\newblock {\em {J}ournal of {E}conometrics}, 189(2):335--345.

\bibitem[Darsow et al.(1992)]{darsow1992}
Darsow, W.~F., Nguyen, B., and Olsen, E.~T. (1992).
\newblock {C}opulas and {M}arkov processes.
\newblock {\em Illinois Journal of Mathematics}, 36(4):600--642.

\bibitem[De~Lira~Salvatierra and Patton(2015)]{salvatierra2015}
De~Lira~Salvatierra, I. and Patton, A.~J. (2015).
\newblock {D}ynamic copula models and high frequency data.
\newblock {\em Journal of Empirical Finance}, 30:120--135.

\bibitem[Demarta and McNeil(2005)]{demarta2005t}
Demarta, S. and McNeil, A.~J. (2005).
\newblock {T}he t copula and related copulas.
\newblock {\em {I}nternational {S}tatistical {R}eview/{R}evue {I}nternationale
  de {S}tatistique}, pages 111--129.

\bibitem[Domma et~al.(2009)]{Domma2009}
Domma, F., Giordano, S., and Perri, P.~F. (2009).
\newblock Statistical modeling of temporal dependence in financial data via a
  copula function.
\newblock {\em {C}ommunications in {S}tatistics - {S}imulation and
  {C}omputation}, 38(4):703--728.

\bibitem[Engle(2002)]{engle2002}
Engle, R. (2002).
\newblock {D}ynamic conditional correlation: {A} simple class of multivariate
  generalized autoregressive conditional heteroskedasticity models.
\newblock {\em {J}ournal of {B}usiness \& {E}conomic {S}tatistics},
  20(3):339--350.

\bibitem[Engle and Kroner(1995)]{engle1995}
Engle, R.~F. and Kroner, K.~F. (1995).
\newblock {M}ultivariate simultaneous generalized {ARCH}.
\newblock {\em {E}conometric {T}heory}, 11(01):122--150.

\bibitem[Fortin and Kuzmics(2002)]{fortin2002}
Fortin, I. and Kuzmics, C. (2002).
\newblock Tail-dependence in stock-return pairs.
\newblock {\em Intelligent Systems in Accounting, Finance and Management},
  11(2):89--107.

\bibitem[Hafner and Manner(2012)]{hafner2012}
Hafner, C.~M. and Manner, H. (2012).
\newblock {D}ynamic stochastic copula models: {E}stimation, inference and
  applications.
\newblock {\em {J}ournal of {A}pplied {E}conometrics}, 27(2):269--295.

\bibitem[Hamao et~al.(1990)]{hamao1990}
Hamao, Y., Masulis, R.~W., and Ng, V. (1990).
\newblock {C}orrelations in price changes and volatility across international
  stock markets.
\newblock {\em {R}eview of {F}inancial studies}, 3(2):281--307.

\bibitem[Hansen and Lunde(2005)]{hansen2005}
Hansen, P.~R. and Lunde, A. (2005).
\newblock A forecast comparison of volatility models: does anything beat a
  {GARCH}(1,1)?
\newblock {\em {J}ournal of {A}pplied {E}conometrics}, 20(7):873--889.

\bibitem[Ibragimov(2009)]{ibragimov2009}
Ibragimov, R. (2009).
\newblock Copula-based characterizations for higher order markov processes.
\newblock {\em {E}conometric {T}heory}, 25(03):819--846.

\bibitem[Joe(1997)]{joe1997}
Joe, H. (1997).
\newblock {\em Multivariate models and multivariate dependence concepts}.
\newblock CRC Press.

\bibitem[Joe(2014)]{Joe2014}
Joe, H. (2014).
\newblock {\em Dependence Modeling with Copulas}.
\newblock {C}hapman and {H}all/{C}{R}{C}.

\bibitem[Junker and May(2005)]{junker2005}
Junker, M. and May, A. (2005).
\newblock Measurement of aggregate risk with copulas.
\newblock {\em The Econometrics Journal}, 8(3):428--454.

\bibitem[{L}ambert and {V}andenhende(2002)]{Lambert2002}
{L}ambert, P. and {V}andenhende, F. (2002).
\newblock {A} copula-based model for multivariate non-normal longitudinal data:
  analysis of a dose titration safety study on a new antidepressant.
\newblock {\em {S}tatistics in {M}edicine}, 21(21):3197--3217.

\bibitem[Min and Czado(2010)]{MinCzado2010}
Min, A. and Czado, C. (2010).
\newblock {B}ayesian inference for multivariate copulas using pair-copula
  constructions.
\newblock {\em Journal of Financial Econometrics}, 8(4):511--546.

\bibitem[Nakatani and Ter{\"a}svirta(2009)]{nakatani2009}
Nakatani, T. and Ter{\"a}svirta, T. (2009).
\newblock {T}esting for volatility interactions in the constant conditional
  correlation {GARCH} model.
\newblock {\em {T}he {E}conometrics {J}ournal}, 12(1):147--163.

\bibitem[Nelsen(2006)]{Nelsen2006}
Nelsen, R.~B. (2006).
\newblock {\em An Introduction to Copulas (Springer Series in Statistics)}.
\newblock {S}pringer-{V}erlag {N}ew {Y}ork, {I}nc., {S}ecaucus, {N}{J},
  {U}{S}{A}.

\bibitem[Oh and Patton(2016a)]{oh2013}
Oh, D.~H. and Patton, A.~J. (2016a).
\newblock {T}ime-varying systemic risk: evidence from a dynamic copula model of
  {CDS} spreads.
\newblock {\em Journal of Business and Economic Statistics}, forthcoming.

\bibitem[Oh and Patton(2016b)]{oh2016}
Oh, D.~H. and Patton, A.~J. (2016b).
\newblock High-dimensional copula-based distributions with mixed frequency
  data.
\newblock {\em Journal of Econometrics}, forthcoming.

\bibitem[Patton(2006)]{Patton2006}
Patton, A.~J. (2006).
\newblock {M}odelling asymmetric exchange rate dependence.
\newblock {\em {I}nternational {E}conomic {R}eview}, 47(2):527--556.

\bibitem[Patton(2012)]{patton2012}
Patton, A.~J. (2012).
\newblock {A} review of copula models for economic time series.
\newblock {\em {J}ournal of {M}ultivariate {A}nalysis}, 110:4--18.

\bibitem[R{\'e}millard et~al.(2012)]{remillard2012}
R{\'e}millard, B., Papageorgiou, N., and Soustra, F. (2012).
\newblock {C}opula-based semiparametric models for multivariate time series.
\newblock {\em {J}ournal of {M}ultivariate {A}nalysis}, 110:30--42.

\bibitem[Roberts and Rosenthal(2009)]{roberts2009}
Roberts, G.~O. and Rosenthal, J.~S. (2009).
\newblock {E}xamples of adaptive {MCMC}.
\newblock {\em {J}ournal of {C}omputational and {G}raphical {S}tatistics},
  18(2):349--367.

\bibitem[Shimazaki and Shinomoto(2010)]{Shimazaki2010}
Shimazaki, H. and Shinomoto, S. (2010).
\newblock {K}ernel bandwidth optimization in spike rate estimation.
\newblock {\em {J}. {C}omput. {N}eurosci.}, 29(1-2):171--182.

\bibitem[Sklar(1959)]{Sklar1959}
Sklar, A. (1959).
\newblock {\em {F}onctions de {R}{\'e}partition {\`{A}} {N} {D}imensions {E}t
  {L}eurs {M}arges}.
\newblock {U}niversit{\'e} {P}aris 8.

\bibitem[Smith et~al.(2010)]{SmithMinAlmCza2010}
Smith, M., Min, A., Almeida, C., and Czado, C. (2010).
\newblock Modeling longitudinal data using a pair-copula decomposition of
  serial dependence.
\newblock {\em {J}ournal of the {A}merican {S}tatistical {A}ssociation},
  105(492):1467--1479.

\bibitem[Smith(2015)]{Smith2015}
Smith, M.~S. (2015).
\newblock Copula modelling of dependence in multivariate time series.
\newblock {\em {I}nternational {J}ournal of {F}orecasting}, 31(3):815 -- 833.

\bibitem[Smith and Maneesoonthorn(2016)]{SmithMan2016}
Smith, M.~S. and Maneesoonthorn, W. (2016).
\newblock Inversion copulas from nonlinear state space models.
\newblock {\em {W}orking {P}aper}.

\bibitem[Smith and Vahey(2016)]{smithvahey2015}
Smith, M.~S. and Vahey, S.~P. (2016).
\newblock Asymmetric density forecasts for {U}{S} macroeconomic variables from
  a {G}aussian copula model of cross-sectional and serial dependence.
\newblock {\em {J}ournal of {B}usiness and {E}conomic {S}tatistics},
  34(3):416--434.

\end{thebibliography}
